\documentclass{article}

\usepackage{PRIMEarxiv}
\usepackage{fancyhdr}
\usepackage{microtype}

\usepackage[colorlinks,linkcolor=black,citecolor=blue,urlcolor=blue]{hyperref}
\usepackage[dvipsnames,svgnames,x11names]{xcolor}

\usepackage{indentfirst}
\usepackage{microtype}

\usepackage{amsmath}
\usepackage{amssymb}
\usepackage{romannum}
\usepackage[T1]{fontenc}
\usepackage[htt]{hyphenat} 

\usepackage{graphicx}
\usepackage{listings}
\usepackage{multirow}
\usepackage{multicol}
\usepackage{makecell}
\usepackage[figuresright]{rotating}

\usepackage{siunitx} 
\usepackage{booktabs}
\usepackage{adjustbox}
\usepackage{tabularx}
\usepackage{array}
\usepackage[utf8]{inputenc}

\def\todo#1{#1}

\def\api#1{\texttt{#1}}

\title{\api{TBPLaS} 2.0: a Tight-Binding Package for Large-scale Simulation}
\author{
	Yunhai~Li$^{1,2}$, Zewen~Wu$^{1,2}$, Miao~Zhang$^1$, Junyi~Wang$^1$, Shengjun~Yuan$^{1,2,3*}$ \\
    $^1$Quantum Computation Division, Wuhan Institute of Quantum Technology \\
	Wuhan 430206, China\\
    $^2$Key Laboratory of Artificial Micro- and Nano-structures of Ministry of Education and \\School of Physics and Technology, Wuhan University \\
	Wuhan 430072, China\\
	$^3$School of Artificial Intelligence, Wuhan University \\
	Wuhan 430072, China \\
	*E-mail:~\texttt{s.yuan@whu.edu.cn} \\
}

\begin{document}
\maketitle

\begin{abstract}
The common exact diagonalization-based techniques to solving tight-binding models suffer from $\mathcal{O}(N^2)$ and $\mathcal{O}(N^3)$ scaling with respect to model size in memory and CPU time, hindering their applications in large tight-binding models. On the contrary, the tight-binding propagation method (TBPM) can achieve linear scaling in both memory and CPU time, and is capable of handling large tight-binding models with billions of orbitals. In this paper, we introduce version 2.0 of \api{TBPLaS}, a package for large-scale simulation based on TBPM \cite{Li2023}. This new version brings significant improvements with many new features. Existing Python/Cython modeling tools have been \todo{thoroughly optimized}, and a compatible C++ implementation of the modeling tools is now available, offering \todo{efficiency enhancement of several orders}. The solvers have been rewritten in C++ from scratch, with the efficiency \todo{enhanced by several times or even by an order of magnitude}. The workflow of utilizing solvers has also been unified into a more comprehensive and consistent manner. New features include spin texture, Berry curvature and Chern number calculation, partial diagonalization for specific eigenvalues and eigenstates, analytical Hamiltonian, and GPU computing support. The documentation and tutorials have also been updated to the new version. In this paper, we discuss the revisions with respect to version 1.3  and \todo{demonstrate the new features}. Benchmarks on modeling tools and solvers are also provided. \end{abstract}


\footnotetext{Yunhai Li and Zewen Wu contribute equally to this work.}

\setlength{\parindent}{1em}
\section{Introduction}
\label{intro}

Tight-binding (TB) theory \cite{Goringe1997,Slater1954} is a powerful tool in solid state physics, chemistry and materials science. It can not only inspire physical insights via analytical solution to the problem, but also evaluate the physical and chemical properties of large models at a relatively low cost compared with density functional theory (DFT) and wavefunction-based quantum chemistry techniques. The common workflow of utilizing TB theory involves the construction and diagonalization of the Hamiltonian matrix, followed by post-processing the eigenvalues and eigenstates to yield the desired quantities. The memory and CPU time costs of exact diagonalization scale as $\mathcal{O}(N^2)$ and $\mathcal{O}(N^3)$ with respect to model size, which limits its application to models with tens of thousands of orbitals at most. Tight-binding propagation method (TBPM) \cite{Logemann2015,Slotman2015,Yuan2011,Hams2000,Yuan2010}, on the other hand, tackles the eigenvalue problem by introducing the correlation functions, which are determined by the time-dependent wave function. Post-processing the correlation functions yields the same physical quantities as exact diagonalization, but at an ultralow computational cost. By expanding the propagation operator in Chebyshev polynomials and taking advantage of the sparsity of Hamiltonian matrix, linear scaling can be achieved in both memory and CPU time costs with respect to model size. Therefore, TBPM can solve ultra-large models with billions of orbitals. In this respect, we have developed the \api{TBPLaS} (Tight-Binding Package for Large-scale Simulation) package \cite{Li2023}. \api{TBPLaS} implements TBPM as well as exact diagonalization, kernel polynomial method (KPM) and Haydock recursive method. Current capabilities of \api{TBPLaS} include the evaluation of electronic structure including band structure, density of states (DOS), topological properties including $\mathbb{Z}_2$ invariant, response properties including local density of states (LDOS), dynamic polarization, dielectric function, electric (DC) and optical (AC) conductivities, Hall conductivity, etc., as described in the article \cite{Li2023} for version 1.3 of the package. The computationally demanding part of \api{TBPLaS} is written in Cython and FORTRAN, while the user interface is implemented in Python, ensuring both efficiency and user friendliness. Since the first public release in 2022, \api{TBPLaS} has established an international user base exceeding 250 researchers and has been employed in research projects on two-dimensional materials \cite{Wang2024,Ren2023,Wang2022,Liu2022}, Moir\'{e} super lattices \cite{Cui2024,Wu2024,Li2024,Gu2024,Hao2024,Hu2023,Long2023,Meng2023,Kuang2022,Long2022}, fractals \cite{Yao2024,Yang2022} and quasicrystals \cite{Yu2022,Yu2022a}.

\begin{figure}[h]
	\centering
	\includegraphics[width=1.0\linewidth]{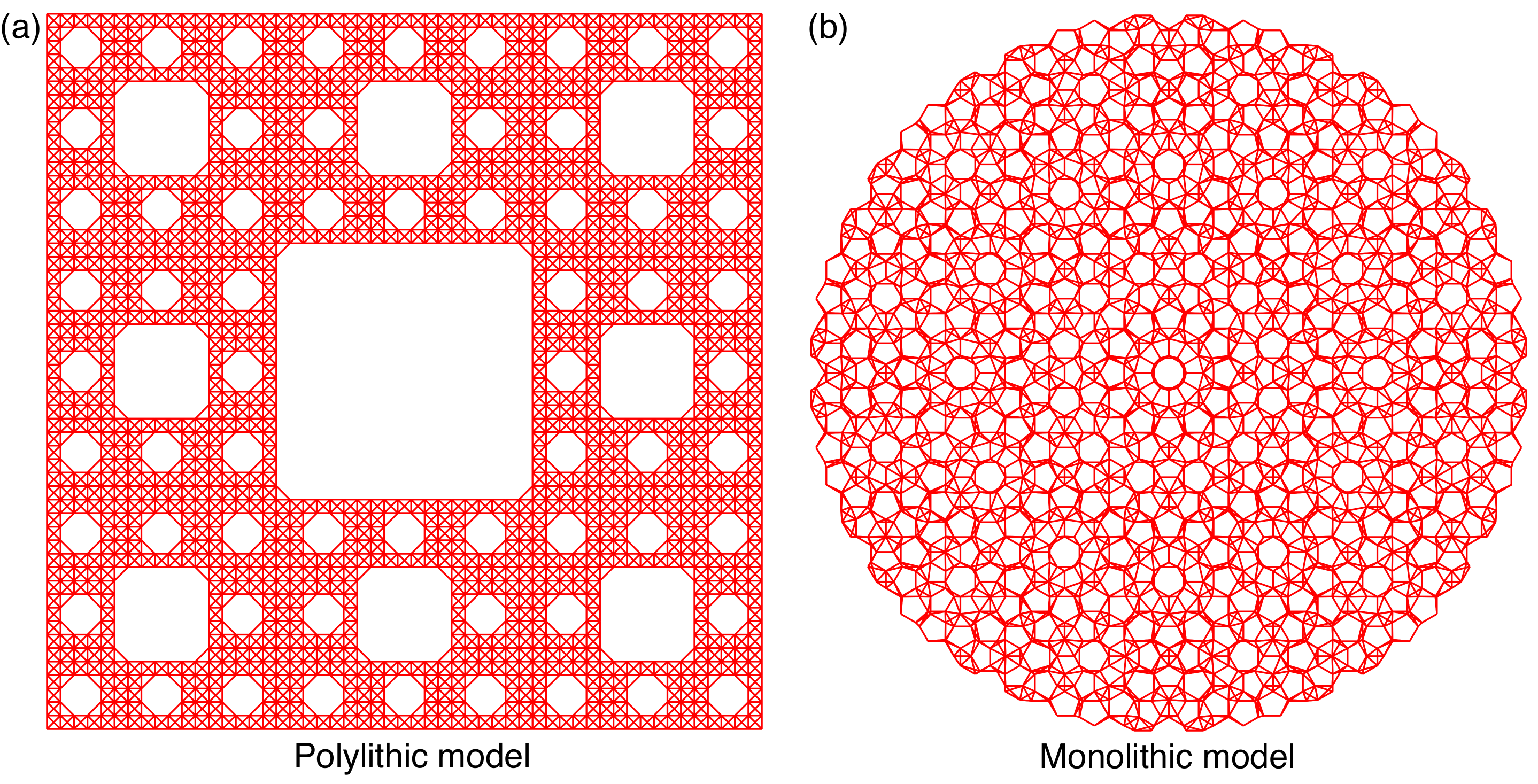}
	\caption{Examples of (a) polylithic and (b) monolithic models.}
	\label{fig:poly_mono}
\end{figure}

Despite the successes, there are still some technical debts to be paid off in both user and developer aspects. Firstly, the modeling tools are not fast enough. Version 1.3 of \api{TBPLaS} provides two categories of modeling tools, namely the Python-based \api{PrimitiveCell} and \api{PCInterHopping}, and Cython-based \api{SuperCell}, \api{SCInterHopping} and \api{Sample} classes. The former is for small and moderate models, while the latter is for large polylithic models that can be formed by replicating the primitive cell following up to given dimension under specific boundary condition. However, there are many monolithic models that cannot be trivially constructed by simply replicating the primitive cell, as shown in Fig. \ref{fig:poly_mono}. For such cases, the user must restore to the Python-based modeling tools, which are slow for large models. The second debt lies in the solvers. Although the computational demanding parts of the solvers are written in Cython and FORTRAN, a significant portion of the source code remains in Python, resulting in slow execution and excessive resource consumption. Another problem is the inconsistencies in the usage of solvers. For the computation of band structure and DOS, the user can call the \api{calc\_bands} and \api{calc\_dos} methods of the model class directly, but for response properties and TBPM algorithms, the user must instantiate the corresponding solvers explicitly. For diagonalization-based solvers, computation parameters are passed as functional arguments. But for TBPM, parameters must be stored in the \api{config} attribute of the solver. Common parameters and outputs are shared across diagonalization and TBPM solvers, but differ in names, units, and default values, causing confusion and steepening the learning curve. The third debt is the build system. Since \api{TBPLaS} uses three programming languages, namely Python, Cython and FORTRAN, additional compilers and build configurations are required, which complicate the build system and cause more compatibility issues. For instance, native build of version 1.3 is impossible on Windows due to compiler incompatibilities.

For developers, the foremost obstacle is the legacy FORTRAN source code. Most subroutines feature lengthy, error-prone parameter lists with inadequate documentation regarding parameter references, input/output array dimensions, and units. Some subroutines even contradict the references, compounding difficulties in maintaining and extending the codebase. Another issue is FORTRAN's ecosystem compared to C and C++. As established industrial languages, C and C++ have a rich set of ready-to-use SDKs, user-friendly IDEs, and a large developer pool. These advantages became evident after several professional software engineers joined the \api{TBPLaS} development team. The C++ language further offers powerful language features such as \textit{move} semantics, smart pointers and template metaprogramming, which can be leveraged to boost software development. Given these concerns, transitioning from FORTRAN to C++ as the primary development language has become essential.

These issues have been resolved in \api{TBPLaS} 2.0. With the codebase increased from 26,800 to 61,200 lines, this new version brings significant improvements and many new features. Existing Python/Cython modeling tools have been \todo{thoroughly optimized for efficiency}. Meanwhile, a compatible C++ implementation of the modeling tools has been provided for advanced users. The solvers have been rewritten in C++ from scratch, with significant efficiency and maintainability enhancement. The workflow of utilizing solvers has been unified into a more comprehensive and consistent manner. New features include spin texture, Berry curvature and Chern number calculation, partial diagonalization for specific eigenvalues and eigenstates, analytical Hamiltonian and GPU computing support for TBPM algorithms. The documentation and tutorials have also been updated to the new version.

The paper is organized as follows. Section \ref{revisions} introduces the updates and new functionalities introduced in version 2.0. Section \ref{usage} provides updated guidance on installation and usage. Section \ref{benchmark} presents performance benchmarks against version 1.3. Finally, Section \ref{summary} discusses the conclusions and future development directions. \section{Revisions}
\label{revisions}

\subsection{Modeling tools}
\subsubsection{Optimization of Python/Cython implementation}
The Python-based modeling tools of version 1.3 incorporate an input validation system for detecting invalid user input. Accordingly, a hierarchy of error classes has been designed to provide detailed debugging messages. For example, the \api{\_check\_hop\_index} method of \api{PrimitiveCell} class is a common utility for verifying the cell index and orbital pair in a hopping term, which should be called by any method manipulating hopping terms, e.g., \api{add\_hopping}.
\begin{lstlisting}[language=python]
class PrimitiveCell(Lockable):
    def _check_hop_index(self, rn, orb_i, orb_j):
        rn, legal = check_coord(rn)
        if not legal:
            raise exc.CellIndexLenError(rn)
        num_orbitals = len(self.orbital_list)
        if not (0 <= orb_i < num_orbitals):
            raise exc.PCOrbIndexError(orb_i)
        if not (0 <= orb_j < num_orbitals):
            raise exc.PCOrbIndexError(orb_j)
        if rn == (0, 0, 0) and orb_i == orb_j:
            raise exc.PCHopDiagonalError(rn, orb_i)
        return rn, orb_i, orb_j
\end{lstlisting}
If any of the preconditions in the \textit{if} statements are violated, then the corresponding errors will be raised, terminating the program and displaying the debugging messages. In this approach, the waste of computational resources is avoided.

The input validation system, however, has its own overhead. Programs typically run in two modes: debug mode for eliminating bugs and release mode for production use. In Python, these modes are controlled by the \textit{-O} optimization flag. Ideally, the validation system should activate only in debug mode and be disabled in release mode. However, this is not feasible because the checks rely on \textit{if} statements, which inevitably consume CPU cycles in release mode. The error class hierarchy also imposes maintenance challenges. These classes require comprehensive unit tests and up-to-date documentation, both labor-intensive tasks.

The key idea to solve these problems is to distinguish between \textit{bugs} and \textit{exceptions}. Detailed discussion of these concepts is beyond the scope of this paper. Briefly, \textit{bugs} are unintended internal flaws which must be eliminated during development, while \textit{exceptions} are unavoidable external disruptions which need to be properly handled at runtime. Violations of preconditions are unequivocally \textit{bugs}, as they indicate errors in the program logic.

The recommended approach to detect violations of preconditions in Python is through \textit{assert} or the builtin \textit{\_\_debug\_\_} constant, which are active only in debug mode and deactivated automatically in release mode. In version 2.0, the input validation system has been rewritten using this approach. For example, the \api{\_check\_hop\_index} method is now defined as
\begin{lstlisting}[language=python]
class PrimitiveCell(Lockable):
    def _check_hop_index(self,
        rn: rn3_type,
        orb_i: int,
        orb_j: int) -> None:
        num_orb = self.num_orb
        assert 0 <= orb_i < num_orb, f"Orb_i {orb_i} out of range(0, {num_orb})"
        assert 0 <= orb_j < num_orb, f"Orb_j {orb_j} out of range(0, {num_orb})"
        error_msg = f"{rn}{orb_i, orb_i} is a diagonal term"
        assert rn != (0, 0, 0) or orb_i != orb_j, error_msg
\end{lstlisting}
Once the preconditions are violated in debug mode, an \textit{AssertionError} will be raised, carrying the same debugging messages as the error classes in version 1.3. In release mode the validation process is skipped, enhancing the efficiency by more than \todo{30\%}. Since violations are \textit{bugs} rather than \textit{exceptions}, they do not require runtime handling. Consequently, the hierarchy of error classes, exhaustive unit tests, and related documentation are unnecessary. In version 2.0, these error classes are deprecated, significantly reducing maintenance overhead.

The only error class retained in version 2.0 is \api{PCHopNotFoundError} for indicating a missing hopping term. Large primitive cells may contain thousands or millions of hopping terms, making it impossible to keep track of which terms are included in the model and which are not. On the other hand, accessing missing hopping terms is inevitable in some cases, e.g., adding spin-orbit coupling of $\lambda\mathbf{L}\cdot\mathbf{S}$ form, and the situation can be easily recovered once occurred. So we decided to keep the \api{PCHopNotFoundError} error class in version 2.0.

For Cython-based modeling tools, optimization involves the simplification and parallelization of Cython extensions. In version 1.3, the performance critical logic is fully implemented in Cython, which is then converted into C source code and compiled. However, it is difficult to achieve fine-grained control over parallelism in Cython as in native languages like C and C++, due to the global interpreter lock (GIL) and limited language features. Debugging Cython extensions is not an easy task, since the machine-generated C source code is not human-readable. In version 2.0, we have migrated all the core logic to C++ and only use Cython as thin wrapper over C++ stuff. This makes parallelism and debugging much easier. These optimizations have enhanced the efficiency of modeling tools \todo{by several times or even by several orders}. To facilitate easier installation, the C++ source code and Cython wrappers have been consolidated into the \api{tbplas-cpp} package. Further technical details are provided in Section \ref{build_system}.

\subsubsection{New C++ implementation}
\label{rev_model_cpp}
\begin{figure}[h]
	\centering
	\includegraphics[width=1.0\linewidth]{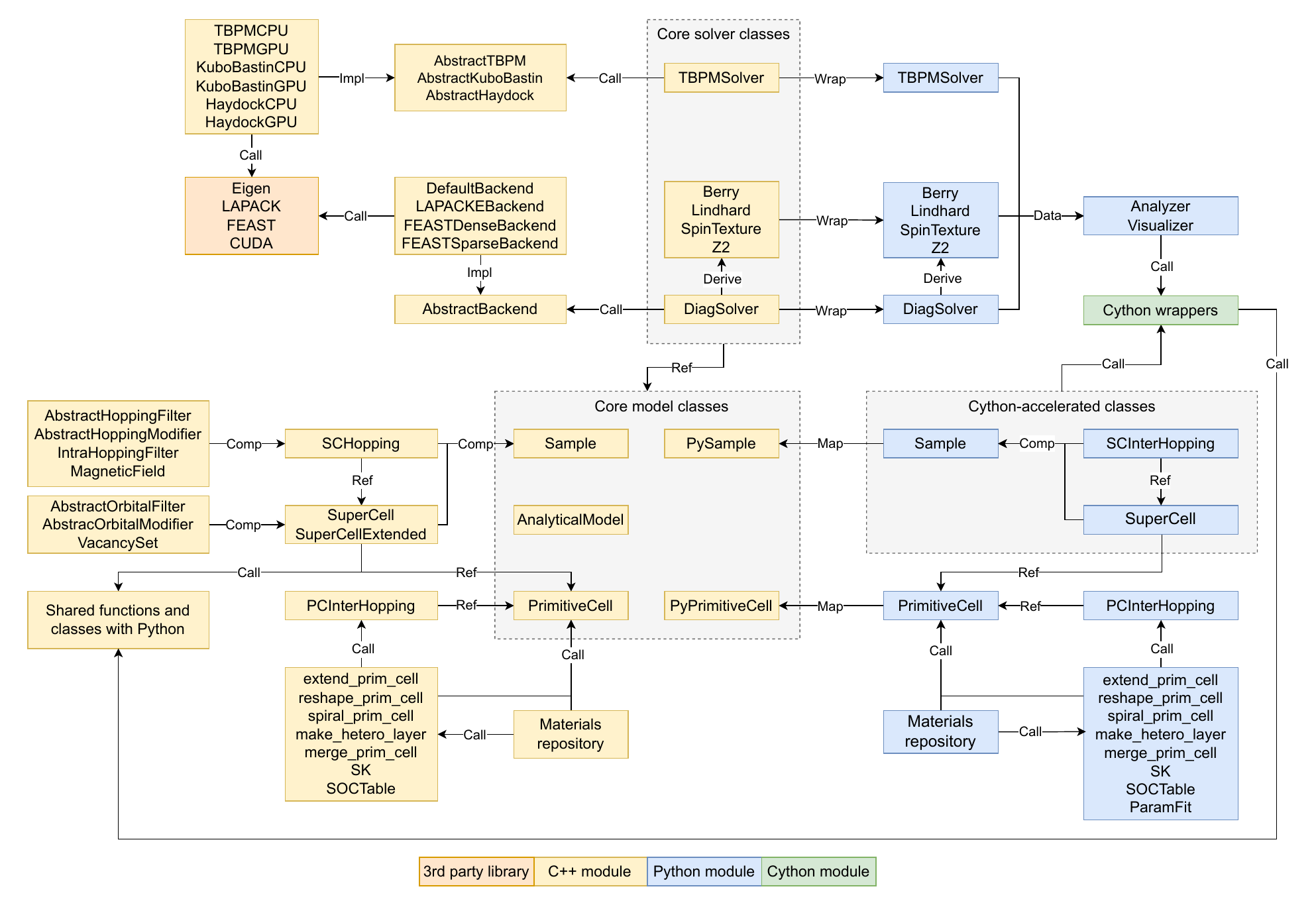}
	\caption{Schematic diagram of classes and functions of \api{TBPLaS} 2.0. $A\xrightarrow{ref}B$ means class A holds a pointer to an instance of class B, while \textit{comp} means A is a component of B. The term \textit{impl} indicates class B is an abstract interface and is implemented by class A. A C++ class can be \textit{wrapped} as a Python class by adding proper methods, while a Python class can be \textit{mapped} to a C++ class by sharing memory, both with the help of Cython wrappers (not shown for clarity).}
	\label{fig:classes}
\end{figure}

Version 2.0 of \api{TBPLaS} brings a brand-new C++ implementation of the modeling tools. The aims are to provide a highly efficient solution in case the Python/Cython-based modeling tools are slow, e.g., when dealing with large monolithic models, and to facilitate incorporation of \api{TBPLaS} into other performance-critical scientific programs. The schematic diagram of the C++ modeling tools and their relation to Python/Cython counterparts is shown in Fig. \ref{fig:classes}. Both the Python-based components (\api{PrimitiveCell}, \api{PCInterHopping}, utilities such as \api{extend\_prim\_cell} and \api{SK}, materials repository) and Cython-based components (\api{SuperCell}, \api{SCInterHopping}, \api{Sample}) have been ported to C++. Unlike the solvers where the Python implementation are wrappers over C++ core, the C++ and Python/Cython implementations of modeling tools are mutually independent sharing a few core functions and a compatible API. The reason is that intensively calling C++ functions from Python, which is inevitable when constructing models, causes significant performance overhead. If Python version of modeling tools were restructured as wrappers over C++ version, both the efficiency of C++ and the flexibility of Python would be lost.

The usage of C++ \api{PrimitiveCell} class is the same as that of Python/Cython version. Users should create an empty primitive cell by providing lattice vectors, which can be either specified manually or generated with the \api{gen\_lattice\_vectors} function. Orbitals and hopping terms should then be added to the primitive cell by calling the \api{add\_orbital} and \api{add\_hopping\ methods} (functions), respectively. Auxiliary classes and functions can be utilized to evaluate on-site energies and hopping terms, e.g., \api{SK} for Slater-Koster formulation, \api{SOCTable} for spin-orbital coupling of $\lambda\mathbf{L}\cdot\mathbf{S}$ type, and \api{find\_neighbors} for identifying neighboring orbitals within cutoff distance. The materials repository contains a set of pre-defined primitive cells available for import. Once the primitive cell is configured, complex models can be constructed using functions like \api{extend\_prim\_cell}, \api{reshape\_prim\_cell}, and \api{merge\_prim\_cell}. Users are advised to consult the article \cite{Li2023} for version 1.3 for detailed description of these tools. Examples demonstrating the usage of the C++ \api{PrimitiveCell} class can be found in Section \ref{basic_modeling} and \ref{advanced_modeling_prim_cell}.

The C++ \api{SuperCell} class is a generalized version of its Python/Cython counterpart, which is limited to polylithic models formed by extending the primitive cell along crystallographic $a$, $b$, and $c$ directions. The C++ version, on the other hand, supports extending the primitive cell along arbitrary directions, similar to the \api{reshape\_prim\_cell} function. This improvement makes constructing twisted hetero-structures much simpler. The second improvement is the approach to handle intra-supercell and inter-supercell hopping terms. In the Python/Cython version, the former are handled by the \api{SuperCell} class itself, while the latter are maintained by the \api{SCInterHopping} class, although they have much in common. In the C++ version, we generalize the container of hopping terms to the \api{SCHopping} class that can handle intra-supercell and inter-supercell hopping terms on the same footing. This consolidation enhances the consistency and usability of modeling tools.

\begin{figure}[h]
	\centering
	\includegraphics[width=0.7\linewidth]{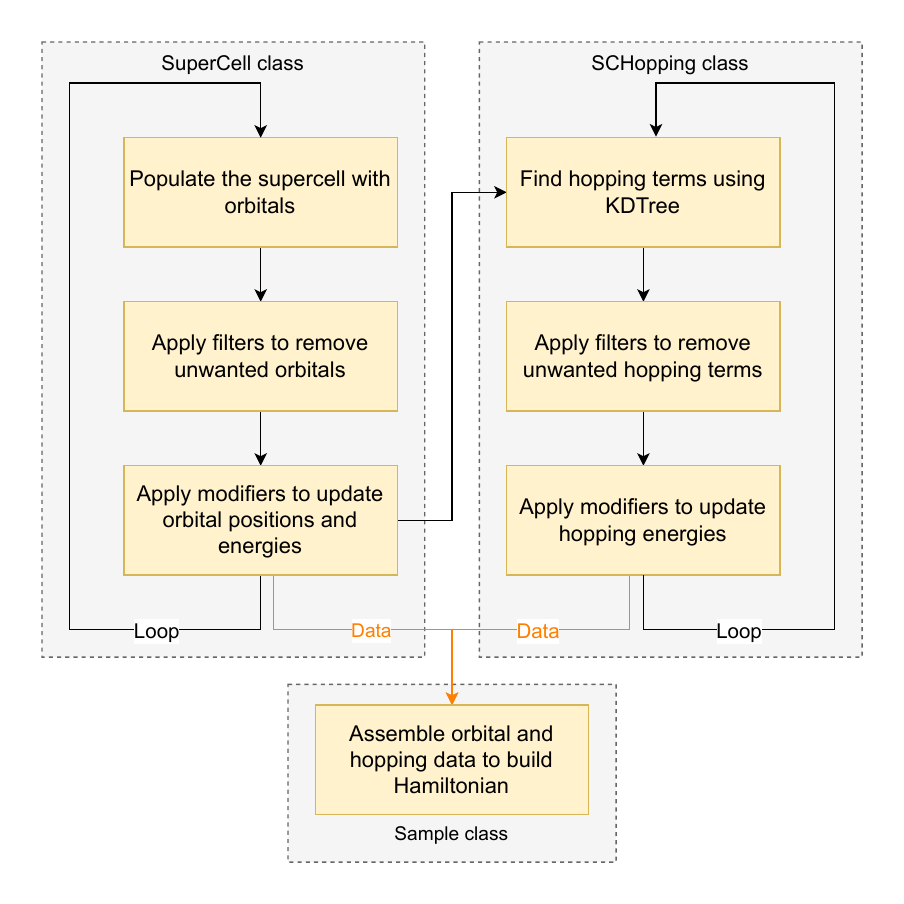}
	\caption{Schematic diagram of the \textit{filter-modifier} pattern for C++ \api{SuperCell}, \api{SCHopping} and \api{Sample} classes. The loops run over all \api{SuperCell} or \api{SCHopping} instances assigned to the sample.}
	\label{fig:pattern}
\end{figure}

Another notable improvement is the approach to manipulating the orbitals and hopping terms. In the Python/Cython version, removal of orbitals should be implemented via vacancies, and modification of orbital positions should be implemented using position modifiers. Orbital energies and hopping terms should be modified by directly changing the array attributes of the \api{Sample} class. The C++ version removes these discrepancies by introducing a unified \textit{filter-modifier} pattern as shown in Fig. \ref{fig:pattern}. Orbitals are first generated by populating the supercell, then filtered by the filters to remove unwanted ones. Finally, positions and energies are modified by the modifiers. We provide two abstract base classes \api{AbstractOrbitalFilter} and \api{AbstractOrbitalModifier}. Users should implement customized filters and modifiers as derived classes of the base classes. For hopping terms, the case is similar, where the customized filters and modifiers should be derived from \api{AbstractHoppingFilter} and \api{AbstractHoppingModifier}. The \textit{filter-modifier} pattern offers a unified and comprehensible approach to implementing perturbations like vacancies, strains, electric and magnetic fields, etc. Demonstration of this pattern can be found in Section \ref{advanced_modeling_sample}.

Finally, we discuss the compatibility and efficiency of C++ version of modeling tools. The \api{PrimitiveCell} class and relevant modeling tools have good compatibility with the Python/Cython counterparts, since they share the same API. The incompatibilities mainly arise from the different semantics of C++ and Python, e.g., C++ lacks the flexibility of keyword arguments and memory safety of garbage collection of Python. For the \api{SuperCell}, \api{SCHopping}, and \api{Sample} classes, although the C++ version uses generalized algorithms and new workflow, the legacy workflow is still available. For example, the C++ \api{SuperCellExtended} class works similarly to the Python/Cython \api{SuperCell} class. In fact, they share core functionality as shown in Fig. \ref{fig:classes}. In addition to the \textit{filter-modifier} pattern, perturbations can also be implemented by modifying the array attributes of the C++ \api{Sample} class. For example, both the \api{MagneticField} hopping modifier and the \api{apply\_magnetic\_field} method of Sample class can impose a perpendicular magnetic field via Peierls substitution \cite{vonsovsky1989quantum}. Regarding efficiency, according to our tests on twisted-bilayer graphene, quasicrystals and fractals, the C++ version of modeling tools is \todo{an order of magnitude faster} than the Python/Cython version in most cases. More details are provided in Section \ref{benchmark_model}.

\subsection{Solvers}
\subsubsection{Migration to C++}
\label{rev_solver_cpp}
The diagonalization and TBPM solvers of version 1.3 feature a mixed Python/FORTRAN architecture. The main parts are written in Python, whereas performance-critical parts such as diagonalization and post-processing are implemented in FORTRAN. In other words, they are \textit{Python solvers with FORTRAN extensions}. This architecture, while retaining efficiency and flexibility, has its own disadvantages. A significant portion of the source codes are written in Python, reducing the efficiency of the solvers. Due to the significant performance overhead of intensive cross-language function calls, eigenvalues and eigenstates for all $\mathbf{k}$-points must be computed simultaneously in Python before being passed to FORTRAN subroutines, which leads to excessive memory consumption. Additionally, since the solvers are mainly written in Python, it is difficult to integrate them into other performance-critical scientific applications developed entirely in compiled languages, imposing limitations on the application of \api{TBPLaS}.

These problems have been solved in version 2.0, with all the solvers rewritten from scratch.  As shown in Fig. \ref{fig:classes}, all the logic has been migrated to C++. The C++ solver classes are fully functional and can be directly integrated into high-performance scientific applications, while the Python solvers are now merely \textit{wrappers over C++ core}. Data exchange between the C++ core and Python wrappers is achieved with file-based io and shared memory. The C++ solver classes make extensive use of object-oriented programming (OOP) and template-based metaprogramming. For example, the \api{DiagSolver} class is a base class implementing diagonalization methods, while \api{Berry}, \api{Lindhard}, \api{SpinTexture} and \api{Z2} inherit from it and extend its functionality. Both the diagonalization and TBPM solvers take the model class as template argument and hold a pointer to the model, making them applicable to any model class that implements the required methods, e.g. user-defined models as derived class of \api{AnalyticalModel}.

To achieve run-time switching between different math library vendors and computing devices, we employ the \textit{pointer to implementation} (PIMPL) pattern. A virtual interface class is defined and declares the abstract methods that must be implemented by derived classes. Subsequently, implementation classes define these methods and handle the technical details of interacting with specific math libraries. In this approach, superior flexibility and extensibility can be achieved. For example, the \api{TBPMGPU} class implements the \api{AbstractTBPM} interface class, enabling switching to GPU as the computing device at run-time by simply changing the \api{config.algo} attribute of \api{TBPMSolver} instance. And support for new math libraries can be easily added by introducing new implementations of the interface class. These design patterns maximize code reuse and modularity, significantly reducing development and testing efforts.

In addition to refactoring the high-level architecture, the mathematical subroutines under the hood have also been rewritten from scratch. We have carefully examined the mathematical formulae of TBPM algorithms and have introduced many composite functions that avoid the use of temporary arrays and unnecessary copy assignments. The algorithms themselves have also been thoroughly optimized. For example,  in the evaluation of Hall conductivity \cite{Garcia2015} we need to act the Hamiltonian and current operator on the wave functions. In version 1.3 it is implemented as
\begin{lstlisting}[language=fortran]
DO j = 3, n_kernel
    wf_DimKern(:, j) = H_csr * wf_DimKern(:, j-1)
    CALL axpby(-1D0, wf_DimKern(:, j-2), 2D0, wf_DimKern(:, j))
END DO
DO j = 1, n_kernel
    wf0 = copy(wf_DimKern(:, j))
    wf_DimKern(:, j) = cur_csr_x * wf0
END DO
\end{lstlisting}
while in version 2.0 the sparse matrix-vector multiplication and \api{axpby} ($y=ax+by$) have been merged into a single call to \api{amxsy} ($y=aMx-y$) with $a$ and $b$ being scalars, $x$ and $y$ being vectors and $M$ being a sparse matrix. The action of current operator on the wave function has also been simplified to the call to \api{mv}. With the help of composite functions, the use of temporary arrays is avoided, reducing the memory access by 50\%.
\begin{lstlisting}[language=c++]
for (int i = 2; i < num_kernel; ++i) {
    p2 = p0;
    h_sparse.amxsy(2.0, *p1, *p0);
    curr_beta->mv(*p2, wf_vb_tn[i]);
    p0 = p1;
    p1 = p2;
}
\end{lstlisting}
Another example is the evaluation of time-dependent wavefunction which requires the summation over Chebyshev series. In version 1.3 it is implemented as
\begin{lstlisting}[language=fortran]
DO i = 4, SIZE(Bes)
    p2 => p0
    CALL amxpy(-2*img_dt, H_csr, p1, p0)
    CALL axpy(2*Bes(i), p2, wf_out)
    p0 => p1
    p1 => p2
END DO
\end{lstlisting}
where \api{amxpy} is defined as $y=aMx+y$ and \api{axpy} is defined as $y=ax+y$. In version 2.0, the imaginary factor \api{img\_dt} has been merged into the Chebyshev coefficients, and the call to \api{amxpy} becomes \api{amxsy} with $a$ being a real number
\begin{lstlisting}[language=c++]
for (size_t n = 3; n < num_series; ++n) {
    p2 = p0;
    h_sparse->amxsy(2.0, *p1, *p0);
    axpy(coeff[n], *p2, wf_out);
    p0 = p1;
    p1 = p2;
}
\end{lstlisting}
Since the sparse matrix $M$ has more non-zero elements than the vector $y$, moving the imaginary factor from \api{amxpy} to \api{axpy} can significantly boost the calculations. Suppose the length of vector $y$ is $N_o$ and each row of sparse matrix $M$ has $N_t$ non-zero elements, then the speed up of C++ implementation can be estimated from the amount of float number multiplications as
\begin{equation}
\frac{4N_oN_t+2N_o}{2N_oN_t+4N_o} = \frac{2N_t+1}{N_t+2}
\end{equation}
For monolayer graphene $N_t=3$, leading to a speed up of 40\%. 
Owing to the optimization, the solvers of version 2.0 are much more efficient than that of version 1.3. According to our benchmarks, most of the capabilities of diagonalization and TBPM solvers are now \todo{several times faster}. The DC conductivity, Hall conductivity and Haydock recursive method for LDOS are even \todo{an order of magnitude faster} than the 1.3 version. Detailed discussions on the benchmarks can be found in Section \ref{benchmark_solver}.

\subsubsection{Unified workflow}
\label{rev_solver_workflow}
\begin{figure}[h]
	\centering
	\includegraphics[width=0.5\linewidth]{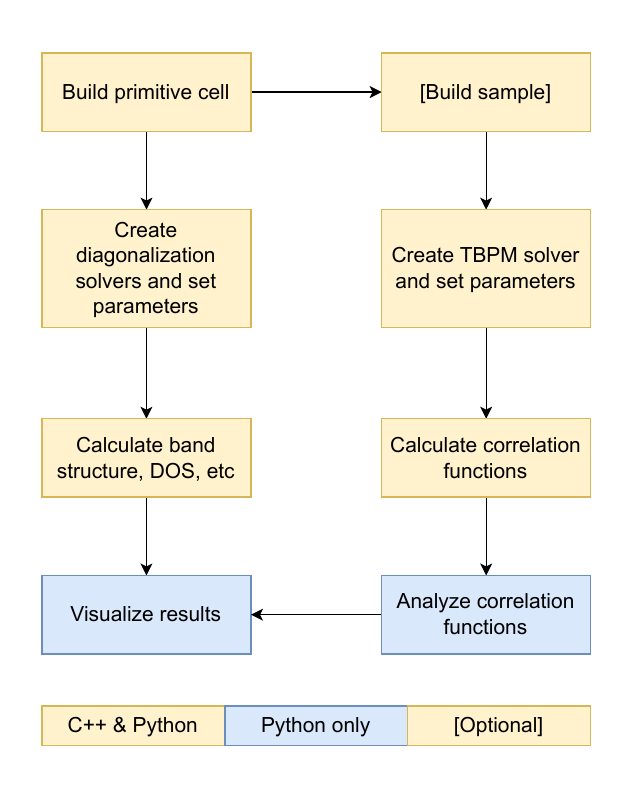}
	\caption{Workflow of usage of \api{TBPLaS} 2.0. Procedures in yellow rectangles are applicable to both Python and C++ APIs, while those in blue rectangles must be done with the Python API. Square brackets indicate that the procedures are optional and can be skipped when using the C++ API.}
	\label{fig:workflow}
\end{figure}

In version 2.0, the workflow has been unified into a more comprehensive and consistent manner. As shown in Fig. \ref{fig:workflow}, the workflow also begins with constructing the model from either \api{PrimitiveCell} or \api{Sample} classes depending on the model size and calculation type, similar to version 1.3. The difference is that the use of \api{Sample} class is optional when using the C++ API for TBPM calculations,  since the \api{PrimitiveCell} class is already efficient enough. Another difference is that the \api{Sample} class is exclusively for TBPM calculations in version 2.0 for both Python and C++ APIs, since it is dedicated to extra-large models which are far beyond the capabilities of diagonalization-based methods. Then diagonalization or TBPM solvers are created from the model and calculation parameters are set. Unlike in version 1.3, where the model classes generate the band structure and DOS solvers implicitly, all diagonalization-based solvers must be explicitly instantiated in version 2.0. In other words, the \api{calc\_bands} and \api{calc\_dos} methods of \api{PrimitiveCell} and \api{Sample} classes have been removed. Also, the parameters should be specified via the built-in \api{config} attribute of the solvers for both diagonalization and TBPM in version 2.0. The aim of these changes is to resolve the ambiguities and inconsistencies in version 1.3. Finally, the proper methods of the solvers are called to evaluate the desired properties, which are which are then post-processed and visualized.

Most of the procedures are applicable to both Python and C++ APIs, with the exceptions of post-processing and visualization, which are exclusive to the Python API. A set of I/O functions have been implemented to load the data files produced by C++ backends and integrate seamlessly with the Python post-processing and visualization procedures. Examples on the workflow can be found in Section \ref{workflow}.

\subsection{New features}
\subsubsection{Spin texture}
Spin texture refers to the expectation values of the Pauli operators $\hat{\sigma}_I$ as the function $\mathbf{k}$-point in the basis of eigenstates ${\psi_{n\mathbf{k}}}$, with $I\in x,y,z$ and $n$ being the band index. For models with non-zero spin-orbital coupling (SOC), $\hat{\sigma}_z$ is no longer conserved and the spin-texture becomes non-trivial. \api{TBPLaS} 2.0 implements the \api{SpinTexture} Python and C++ solver classes for evaluating the spin texture and spin-projected band structure. The spin texture is calculated as
\begin{equation}
    S_{I,n}(\mathbf{k}) = \langle\psi_{n\mathbf{k}}\vert\hat{\sigma}_I\vert\psi_{n\mathbf{k}}\rangle = \sum_{ij\alpha\beta}C_{n\mathbf{k},i\alpha}^{*}\sigma_{I,\alpha\beta}C_{n\mathbf{k},j\beta}
\end{equation}
where $C_{n\mathbf{k}}$ is the coefficients of $n$-th eigenstate at $\mathbf{k}$-point, $i,j$ are the orbital indices and $\alpha,\beta$ denote the spin channels ($\uparrow\downarrow$). Accordingly, the \api{Visualizer} class has two new methods \api{plot\_scalar} and \api{plot\_vector} for plotting $S_{z,n}$ as scalar field and $(S_{x,n}, S_{y,n})$ as vector field of $\mathbf{k}$-point, respectively. Contour plot of spin texture within specific energy range is also supported.

\subsubsection{Berry curvature and Chern number}
\api{TBPLaS} 2.0 implements the \api{Berry} Python and C++ solver classes for calculating the Berry curvature and topological Chern number. Both the Kubo formula and the Wilson loop methods have been implemented. With Kubo formula, the Berry curvature for each band is evaluated as
\begin{equation}
\Omega _{xy}^{n}\left( \mathbf{k} \right) =-2\mathrm{Im}\sum_{m\ne \mathrm{n}}{\frac{\langle {u_n}_{\mathbf{k}}|\frac{\partial H}{\partial \mathbf{k}_x}|{u_m}_{\mathbf{k}}\rangle \langle {u_m}_{\mathbf{k}}|\frac{\partial H}{\partial \mathbf{k}_y}|{u_n}_{\mathbf{k}}\rangle}{\left( E_{m\mathbf{k}}-{E_n}_{\mathbf{k}} \right) ^2}}
\end{equation}
with $u_{n\mathbf{k}}$ and $u_{m\mathbf{k}}$ being the periodic parts of Bloch wave functions and also the eigenstates of Hamiltonian $H(\mathbf{k})$ in convention I (atomic gauge) \cite{coh_python_2022}. $E_{m\mathbf{k}}$ and $E_{n\mathbf{k}}$ are the eigenvalues. In the Wilson loop method, the total Berry curvature is evaluated by considering the local Berry phase on the loop around a small plaquette with vertices ${\{\mathbf{k}_i}\}$
\begin{equation}
\phi _{\mathrm{local}} =-\mathrm{arg}\prod_i{\det M^{\mathbf{k}_i,\mathbf{k}_{i+1}}}
\end{equation}
where $M_{nm}^{\mathbf{k}_i,\mathbf{k}_{i+1}}$ is defined as
\begin{equation}
\label{eq:unk_ovelap}
M_{nm}^{\mathbf{k}_i,\mathbf{k}_{i+1}}=\langle {u_n}_{\mathbf{k}_i}|{u_m}_{\mathbf{k}_{i+1}}\rangle =\sum_j{C_{n\mathbf{k}_i,j}^{*}C_{m\mathbf{k}_{i+1},j}}
\end{equation}
Then the Berry curvature can be determined as
\begin{equation}
\Omega _{xy}\left( \mathbf{k} \right) =\frac{\phi _{\mathrm{local}}}{\mathrm{d}S_z}
\end{equation}
From the Berry curvature we can get the Berry phase $\phi$ and Chern number $c_n$ as
\begin{equation}
\phi =2\pi c_n=\int_{\mathrm{FBZ}}{\Omega _{xy}\left( \mathbf{k} \right) \mathrm{d}S_z}
\end{equation}
The integration is performed on the $xOy$ plane of the first Brillouin zone (FBZ) and $\mathrm{d}S_z$ is the unit area perpendicular to the $z$-axis. Visualization of Berry curvature is similar to spin texture.

\subsubsection{Partial diagonalization}
Support for partial diagonalization has been introduced into \api{TBPLaS} 2.0 based on the FEAST library \cite{feast}, a package for solving various families of eigenvalue problems and addressing the issues of numerical accuracy, robustness, performance and parallel scalability. Owing to FEAST, \api{TBPLaS} 2.0 can search for eigenvalues and eigenstates within a specific energy range using the contour integration algorithm and can handle both dense and sparse Hamiltonian. Fine control over the FEAST library, such as the initial guess of eigenstates, the size of searching subspace and the fpm parameter array is also supported. Diagonalization-based algorithms can also be used to compute band structure, density of states, spin texture, etc., on top of the eigenvalues and eigenstates.

\subsubsection{Analytical Hamiltonian}
\api{TBPLaS} 2.0 supports defining tight-binding models directly from the analytical Hamiltonian formula via the C++ API. User-defined models must inherit the abstract \api{AnalyticalModel} class and implement the specific methods required by the solvers. For example, diagonalization-based solvers require the functions \api{build\_ham\_dense} and \api{build\_ham\_csr} to set up the dense and sparse Hamiltonian. \api{Berry} optionally requires \api{build\_ham\_der\_dense} for evaluating the derivatives of Hamiltonian, while \api{Lindhard} additionally requires \api{build\_ham\_dr\_coo} and \api{build\_density\_coeff} to evaluate the hopping data and density operators, respectively. TBPM solvers require \api{build\_ham\_csr} as well as \api{build\_ham\_curr\_csr} to set up the sparse current operators. To utilize the user-defined model, the model class must be passed as the template argument of the solver class, following the philosophy described in Section \ref{rev_solver_cpp}. The workflow is the same as that of ordinary model classes, e.g., \api{PrimitiveCell} and \api{Sample} as described in Section \ref{rev_solver_workflow}.

\subsubsection{GPU computing}
GPU computing has been implemented in \api{TBPLaS} 2.0 supporting all the TBPM algorithms, the Kubo-Bastin method for Hall conductivity, and the Haydock recursive method. Excellent speed up has been achieved with respect to CPU according to the benchmarks in Section \ref{benchmark_solver}. Run-time switching of computing devices between GPU and CPU is also supported by taking advantage of the PIMPL pattern discussed in Section \ref{rev_solver_cpp}.

\subsection{Miscellaneous}
\subsubsection{Build system}
\label{build_system}
Starting from version 2.0, \api{TBPLaS} will be released as two separate packages, namely \api{tbplas-py} and \api{tbplas-cpp}, which contain the Python and C++ components, respectively. The two packages are loosely coupled through the \api{TBPLAS\_CORE\_PATH} environment variable specifying the installation path of \api{tbplas-cpp}. The aim is to decouple the Python interface from the C++ core. Version 2.0 brings many new features, and some of them are mutually exclusive. So, it becomes necessary to have a unified Python interface that can be dynamically switched between multiple C++ cores built with different features, e.g., one with CUDA and the other with MPI support. The separation of Python and C++ components also simplifies the installation procedure. \api{tbplas-py} can be installed just as a common Python package from source or via the wheel installer. No configuration or compilation is required. \api{tbplas-cpp} features a CMake-based build system and can be compiled and installed as a common C++ package. The build system has a rich set of configuration options and a dedicated validation procedure for checking incompatible combinations of the options. Most compilers and math libraries, e.g., GNU Compiler Collection (GCC) \cite{gcc}, Clang/LLVM \cite{llvm}, Intel oneAPI \cite{oneapi}, AMD Optimizing C/C++ Compilers (AOCC) \cite{aocc}, Netlib LAPACK \cite{netlib}, OpenBLAS \cite{openblas}, AMD Optimizing CPU Libraries (AOCL) \cite{aocl} and NVIDIA HPC SDK \cite{nvidia} have been tested and fully supported. Native build of \api{tbplas-cpp} on Windows is now possible after the removal of legacy FORTRAN components, and pre-compiled binary installer is available for download. The detailed installation instructions are provided in Section \ref{install}.

\subsubsection{Parallelization}
The parallelization scheme of version 2.0 is the same as that of version 1.3. For diagonalization-based algorithms, parallelization is achieved by distributing k-points over MPI processes. For each k-point, the diagonalization and post-processing are further parallelized over OpenMP threads. For TBPM algorithms, the random initial states are distributed over MPI processes. For each initial state, the propagation is parallelized using OpenMP threads. The users are recommended the article for version 1.3 for detailed discussion on the parallelization scheme. The change in version 2.0 is that MPI-based parallelization must be enabled during the compilation of \api{tbplas-cpp}, similar to the OpenMP-based parallelization. In other words, it can no longer be enabled by setting the \api{enable\_mpi} argument to true during runtime as in version 1.3. This is because the whole logic of solvers has been moved to C++, and the installation of \api{mpi4py} package is non-trivial. See Section \ref{install} for more details. \section{Usage}
\label{usage}

In this section we demonstrate the installation and usage of \api{TBPLaS} 2.0. The source code, documentation and tutorials are available on the homepage \url{www.tbplas.net}. As aforementioned in Section \ref{build_system}, \api{TBPLaS} is now released as two separate packages, namely \api{tbplas-cpp} and \api{tbplas-py}, which contain the C++ and Python components respectively. Both packages need to be installed for full functionality. Precompiled installers are also available. But for optimal performance and full functionality, native build from source code is recommended, and will be discussed in this section.

The Python/Cython implementation of modeling tools of version 2.0 are compatible with version 1.3. Legacy modeling scripts can be run with version 2.0 with minor modification, even for complicated models such as twisted bilayer graphene, fractal and quasicrystal. On the contrary, the C++ implementation of modeling tools is brand new in version 2.0. The solvers have been rewritten from scratch with significant changes, and the workflow has also been updated to a more consistent and comprehensible manner. Therefore, for the usage of version 2.0 we will focus on the \todo{new workflow} and \todo{C++ modeling tools}. Other \todo{new features} such as spin texture, Berry curvature and analytical Hamiltonian will also be demonstrated. It is worth noting that the examples in the \api{tbplas-cpp-VERSION\_CPP/samples/speedtest} (replace \api{VERSION\_CPP} with the actual version number) directory demonstrate full capabilities of version 2.0, in both Python and C++ implementations .

Some technical issues need to be addressed concerning the C++ example programs in this section and in the source code. Since version 2.0 makes extensive use of template meta-programming, the flags for compiling the example programs must be the same as those for compiling \api{TBPLaS} itself. Otherwise, runtime errors are likely to be encountered, a well-known problem referred to as the dynamic link library (DLL) Hell in software development. To eliminate potential errors, it is recommended to integrate the example programs into \api{TBPLaS} source code as additional build targets. A practical implementation of this strategy can be found in the \api{samples/demo} directory of \api{tbplas-cpp}.

\subsection{Installation}
\label{install}
\subsubsection{Dependencies}

The dependencies of \api{TBPLaS} 2.0 are summarized in Table. \ref{tab:depend}. C++ compiler supporting C++17 standard and OpenMP 4.0 specifications, CMake, Python interpreter and NumPy, SciPy, Matplotlib, Cython, setuptools and build packages are required. Specific features may have additional dependencies. For example, MPI-based parallelization requires a functional MPI implementation, while GPU computing requires either NVIDIA CUDA toolkit or HPC SDK. LAPACK and sparse matrix libraries are required for efficient linear algebra operations, with vendor-provided implementations such as Intel oneAPI and AMD AOCL expected to have optimal performance on their own CPUs. Searching for eigenvalues within a specific energy range requires the FEAST library to be installed, while binary I/O operations need HDF5. LAMMPS and DeepH-pack interfaces require the ASE and h5py Python packages, respectively. The dependencies can be installed from software repositories or built from source code. We recommend installing Python packages in a separate virtual environment. Refer to \ref{virtual_env} for the preparation of the virtual environment.

To ensure that the installation guide and example programs function correctly, we assume that all the dependencies, should be compiled from source code and installed into the \api{\$HOME/tbplas\_install} directory. For example, HDF5 1.14.2 should be installed into \api{\$HOME/tbplas\_install/hdf5-1.14.2}. This is typically achieved by specifying the installation destination using the \api{\mbox{-}\mbox{-}prefix} or \api{CMAKE\_INSTALL\_PREFIX} options. Note that some dependencies may have their own prefix options, or do not provide any such options at all. Discussion on these cases will be beyond the scope of this paper. The users are recommended to consult the installation guides of these packages for correctly installing them into the target directory. After the installation of the dependencies, especially from source code, some relevant environment variables need to be configured, such that the compiler and CMake can find the headers and libraries. We offer a bash script \api{tools/init.sh} in the \api{tbplas-cpp} package for configuring the environment variables. Refer to \ref{bmod} for the usage of this script.

\begin{table}
	\begin{center}
		\caption{Dependencies of \api{TBPLaS} 2.0, along with the requirements and tested versions.}
		\label{tab:depend}
		\begin{tabular}{cccc}
\hline\hline
            Category & Packages & Requirements & Tested versions \\
			\hline
Compiler & C++ Compiler & \makecell{Supporting C++17 \\ and OpenMP 4.0} & \makecell{GCC 7.5.0 \\ Intel oneAPI 2023.1.0 \\ AMD AOCC 5.0.0} \\
            \hline
Builder & CMake & 3.15 or newer & 3.29 \\
            \hline
\multirow{3}{*}{\makecell{Parallel/GPU \\ computing \\framework \\(optional)}} & CUDA toolkit & & 12.4, 12.8 \\
            \cline{2-4}
                                                                         & HPC SDK                       & & 23.5 \\
            \cline{2-4}
                                                                         & MPI          & & \makecell{MPICH 4.1.2\\Intel oneAPI 2023.1.0} \\
            \hline
\multirow{3}{*}{\makecell{Math libraries\\(optional)}} & LAPACK & \makecell{Built with CBLAS \\and LAPACKE \\ interface} & \makecell{Netlib 3.12.0\\OpenBLAS 0.3.28\\Intel oneAPI 2023.1.0\\AOCL 5.0.0} \\
            \cline{2-4}
            & \makecell{Sparse linear \\algebra library} & \makecell{Supporting CSR \\format} & \makecell{Intel oneAPI 2023.1.0\\AMD AOCL 5.0.0} \\
            \cline{2-4}
            & FEAST & 4.0 or newer & 4.0 \\
            \hline
I/O (optional) & HDF5 & \makecell{Built with C++ \\interface} & 1.14.2 \\
            \hline
\multirow{8}{*}{Python} & Python & 3.7 or newer & 3.12.9 \\
            \cline{2-4}
                                                           & NumPy  &              & 1.26.3, 2.2.4 \\
            \cline{2-4}
                                                           & SciPy  &              & 1.11.4, 1.15.2 \\
            \cline{2-4}
                                                           & Matplotlib &          & 3.8.0, 3.10.0 \\
            \cline{2-4}
                                                           & Cython     &          & 3.0.6, 3.0.11 \\
            \cline{2-4}
                                                           & setuptools & 40.8.0 or newer & 40.8.0 \\
            \cline{2-4}
                                                           & build & 1.2.2 or newer & 1.2.2 \\
            \cline{2-4}
                                                           & ASE (optional) &      & 3.24.0 \\
            \cline{2-4}
                                                           & h5py (optional) &     & 3.12.1 \\
			\hline\hline
		\end{tabular}
	\end{center}
\end{table}

\subsubsection{Installation}
The two packages \api{tbplas-cpp} and \api{tbplas-py} can be installed independently from each other, enabling the decoupling of C++ and Python API. The aim of this design is to have a unified Python frontend that can be dynamically switched between C++ backends with different features, e.g., one with CUDA and the other with MPI support. To compile \api{tbplas-cpp}, create the build directory and change to it
\begin{lstlisting}[language=bash]
# Replace VERSION_CPP with the actual version number
tar -xf tbplas-cpp-VERSION_CPP.tar.bz2
cd tbplas-cpp-VERSION_CPP
test -d build && rm -rf build
mkdir build && cd build
\end{lstlisting}
Then invoke CMake with the following options to configure the build
\begin{lstlisting}[language=bash]
cmake .. \
-DCMAKE_INSTALL_PREFIX=$HOME/tbplas_install/tbplas-cpp-VERSION_CPP \
-DCMAKE_C_COMPILER=gcc \
-DCMAKE_CXX_COMPILER=g++ \
-DCMAKE_BUILD_TYPE=Release \
-DBUILD_SHARED_LIBS=on \
-DBUILD_EXAMPLES=on \
-DBUILD_TESTS=off \
-DBUILD_PYTHON_INTERFACE=on \
-DWITH_OPENMP=on \
-DWITH_MPI=off \
-DWITH_CUDA=off \
-DWITH_FEAST=off \
-DWITH_HDF5=off \
-DEIGEN_BACKEND=default \
-DDIAG_BACKEND=default \
-DTBPM_BACKEND=default
\end{lstlisting}
Interpretation of the options
\begin{itemize}
    \item \api{CMAKE\_INSTALL\_PREFIX}: installation destination
    \item \api{CMAKE\_C\_COMPILER}: C compiler
    \item \api{CMAKE\_CXX\_COMPILER}: C++ compiler
    \item \api{BUILD\_EXAMPLES}: whether to build the example programs
    \item \api{WITH\_OPENMP}: whether to enable OpenMP-based parallelization
    \item \api{WITH\_MPI}: whether to enable MPI-based parallelization
    \item \api{WITH\_CUDA}: whether to enable GPU computation based on CUDA
    \item \api{WITH\_FEAST}: whether to enable interface to FEAST library
    \item \api{EIGEN\_BACKEND}: math library for general linear algebra operations
    \item \api{DIAG\_BACKEND}: math library for Hamiltonian diagonalization
    \item \api{TBPM\_BACKEND}: math library for time propagation
\end{itemize}
The user is recommended to customize the options according to their needs and software environment. For example, setting \api{WITH\_MPI} to \api{on} will enable MPI-based parallelization, while setting \api{DIAG\_BACKEND} to \api{openblas} will utilize OpenBLAS for diagonalization-based calculations. Note that some of the options are mutually exclusive. If the configuration succeeds, proceed with the compilation and installation
\begin{lstlisting}[language=bash]
make -j && make install
\end{lstlisting}
Finally, set up the environment variables by
\begin{lstlisting}[language=bash]
export TBPLAS_CPP_INSTALL_PATH=$HOME/tbplas_install/tbplas-cpp-VERSION_CPP
export TBPLAS_CORE_PATH=$TBPLAS_CPP_INSTALL_PATH/lib
\end{lstlisting}
The first line defines the installation directory of \api{tbplas-cpp}, and the second line sets the location of extensions. Add the settings to \api{\$HOME/.bashrc} to make them permanently effective.

The installation of \api{tbplas-py} is much simpler. Unpack the source code and run pip by 
\begin{lstlisting}[language=bash]
# Replace VERSION_PY with the actual version number
tar -xf tbplas_py-VERSION_PY.tar.bz2
cd tbplas_py
pip install .
\end{lstlisting}
which will install \api{tbplas-py} into the virtual environment. Then run the test suite by
\begin{lstlisting}[language=bash]
cd tests
./run_tests.sh
\end{lstlisting}
The band structure, density of states and many other capabilities of diagonalization-based solvers will be demonstrated by the test suite. If everything goes well, then the installation is successful.

\subsection{Workflow}
\label{workflow}
\subsubsection{Basic modeling}
\label{basic_modeling}
To supplement the usage of solvers, we briefly demonstrate how to build tight-binding models with \api{TBPLaS} 2.0 using both Python and C++ APIs in this section. We show only the essential part of the programs, with the complete programs available in the \api{samples/demo} directory of \api{tbplas-cpp}. The following Python functions are defined to build the model of monolayer graphene from \api{PrimitiveCell} and \api{Sample} classes respectively
\begin{lstlisting}[language=python]
def make_graphene_prim_cell() -> tb.PrimitiveCell:
    """Make graphene primitive cell."""
    # Model parameters
    t = -2.7
    lat = 0.246
    f = 1.0 / 3
    onsite = 0.0

    # Generate lattice vectors
    lat_vec = tb.gen_lattice_vectors(a=lat, b=lat, c=1.0, gamma=60)

    # Create primitive cell
    prim_cell = tb.PrimitiveCell(lat_vec, unit=tb.NM)

    # Add orbitals
    prim_cell.add_orbital((0.0, 0.0, 0.0), energy=onsite, label="C_pz")
    prim_cell.add_orbital((f, f, 0.0), energy=onsite, label="C_pz")

    # Add hopping terms
    prim_cell.add_hopping((0, 0, 0), 0, 1, t)
    prim_cell.add_hopping((1, 0, 0), 1, 0, t)
    prim_cell.add_hopping((0, 1, 0), 1, 0, t)
    return prim_cell


def make_graphene_sample(dim: Tuple[int, int, int]) -> tb.Sample:
    """
    Make graphene sample with specific dimension.
    :param dim: dimension along a, b and c directions
    """
    # Create primitive cell
    prim_cell = make_graphene_prim_cell()

    # Create supercell with specific dimension
    super_cell = tb.SuperCell(prim_cell, dim=dim, pbc=(True, True, False))

    # Assemble supercell to a sample
    sample = tb.Sample(super_cell)
    return sample
\end{lstlisting}
The procedures are identical to that of version 1.3. To build the primitive cell, we first evaluate the Cartesian coordinates of lattice vectors from lattice constants. Then we create an empty model and add the orbitals taking their positions and on-site energies as input. Finally, we add the hopping terms reduced by the conjugate relation $H_{ij}(\mathbf{R})=H_{ji}^{*}(\mathbf{-R})$. From the primitive cell, the sample can be constructed simply by specifying the dimension and boundary condition, as is done in function \api{make\_graphene\_sample}. The equivalent C++ functions are defined as
\begin{lstlisting}[language=c++]
model_t make_graphene_prim_cell()
{
    // Model parameters
    constexpr double t = -2.7;
    constexpr double lat = 0.246;
    constexpr double f = 1.0 / 3;
    constexpr double onsite = 0.0;

    // Generate lattice vectors
    Eigen::Matrix3d lat_vec = gen_lattice_vectors(lat, lat, 1.0, 90.0, 90.0, 60.0);
    Eigen::Vector3d origin(0.0, 0.0, 0.0);

    // Create the primitive cell
    model_t prim_cell(lat_vec, origin, NM);

    // Add orbitals, last 0 for labeling C_pz orbital
    prim_cell.add_orbital(0.0, 0.0, 0.0, onsite, 0);
    prim_cell.add_orbital(f, f, 0.0, onsite, 0);

    // Add hopping terms
    prim_cell.add_hopping(0, 0, 0, 0, 1, t);
    prim_cell.add_hopping(1, 0, 0, 1, 0, t);
    prim_cell.add_hopping(0, 1, 0, 1, 0, t);
    return prim_cell;
}

model_t make_graphene_sample(const std::tuple<int, int, int>& dim)
{
    // Create primitive cell
    model_t prim_cell = make_graphene_prim_cell();

    // Extend primitive cell
    model_t sample = extend_prim_cell(prim_cell, dim);
    return sample;
}
\end{lstlisting}
which are much like the Python counterparts. Note that C++ does not support keyword arguments. Some default parameters in the Python version must be explicitly specified in the C++ version, such as the lattice constants in the \api{gen\_lattice\_vectors} call and the origin parameter in the constructor of \api{PrimitiveCell}. For efficiency, the orbital positions and cell indices are specified as plain double numbers and integers in the C++ version, while in the Python version they take the form of tuples. Since the C++ \api{PrimitiveCell} class is fast enough, we can call \api{extend\_prim\_cell} function directly to make a sample, instead of utilizing the \api{Sampl}e class as is done in the Python version.

\subsubsection{Usage of diagonalization-based solvers}
Now we demonstrate the usage of diagonalization-based solvers. The Python function for calculating band structure is defined as
\begin{lstlisting}[language=python]
def test_diag_bands() -> None:
    """Calculate band structure using diagonalization."""
    # Build the model
    model = make_graphene_prim_cell()

    # Create a solver for the model
    solver = tb.DiagSolver(model)

    # Set up parameters of the solver
    k_points = np.array([
        [0.0, 0.0, 0.0],
        [2./3, 1./3, 0.0],
        [0.5, 0.0, 0.0],
        [0.0, 0.0, 0.0]
    ])
    k_path, k_idx = tb.gen_kpath(k_points, (100, 100, 100))
    solver.config.prefix = "graphene"
    solver.config.k_points = k_path

    # Call 'calc_bands' method of solver to evaluate band structure
    # Data files will be saved automatically.
    timer = tb.Timer()
    timer.tic("bands")
    k_len, bands = solver.calc_bands()
    timer.toc("bands")

    # Report time usage and visualization
    if solver.is_master:
        timer.report_total_time()
        vis = tb.Visualizer()
        vis.plot_bands(k_len, bands, k_idx, ["G", "K", "M", "G"])
\end{lstlisting}
Firstly, we call the \api{make\_graphene\_prim\_cell} function to build the primitive cell. Then we create a solver from the \api{DiagSolver} class and specify the prefix of data files and k-points by modifying the config attribute of the solver. Afterwards, we call the \api{calc\_bands} method of the solver to calculate the band structure and visualize the results using the \api{Visualizer} class. The C++ version of function is defined as
\begin{lstlisting}[language=c++]
void test_diag_bands()
{
    // Build the model
    model_t model = make_graphene_prim_cell();

    // Create a solver for the model
    DiagSolver<model_t> solver(model);

    // Set up parameters of the solver
    Eigen::MatrixX3d k_points {
        { 0.0, 0.0, 0.0 },
        { 2. / 3, 1. / 3, 0.0 },
        { 0.5, 0.0, 0.0 }, { 0.0, 0.0, 0.0 }
    };
    Eigen::Matrix3Xd k_path;
    Eigen::VectorXi k_idx;
    std::tie(k_path, k_idx) = gen_kpath(k_points.transpose(), { 100, 100, 100 });
    solver.config.prefix = "graphene";
    solver.config.k_points = k_path;

    // Call 'calc_bands' method of solver to evaluate band structure
    // Data files will be saved automatically.
    Timer timer;
    timer.tic("bands");
    auto data = solver.calc_bands();
    timer.toc("bands");

    // Report time usage
    if (solver.is_master()) {
        timer.report_total_time();
    }
}
\end{lstlisting}
Note that the Eigen C++ library stores matrices in column-major order, while the NumPy Python library uses row-major order by default. So, the \api{k\_path} matrix takes a transposed form in the C++ version, i.e., $N_{k} \times 3$ in Python and $3 \times N_{k}$ in C++. Since the \api{Visualizer} is available only in the Python API, we need to call the I/O functions to load the data files produced by C++ program, as demonstrated in the following function
\begin{lstlisting}[language=python]
def plot_bands(prefix: str):
    k_len, bands = tb.load_bands(prefix)
    k_idx = np.array([0, 100, 200, 300])
    vis = tb.Visualizer()
    vis.plot_bands(k_len, bands, k_idx, ["G", "K", "M", "G"])
\end{lstlisting}
More examples can be found in the \api{samples/speedtest/plot\_diag.py} script of \api{tbplas-cpp}.

\subsubsection{Usage of TBPM solver}
The usage of TBPM solver is like diagonalization-based solvers
\begin{lstlisting}[language=python]
def test_tbpm_dos():
    """Calculate DOS using TBPM."""
    # Build the model
    model = make_graphene_sample(dim=(512, 512, 1))

    # Create a solver for the model
    solver = tb.TBPMSolver(model)

    # Set up parameters of the solver
    solver.config.prefix = "graphene"
    solver.config.num_random_samples = 1
    solver.config.rescale = 9.0
    solver.config.num_time_steps = 1024

    # Call 'calc_corr_dos' method of solver to evaluate DOS correlation function
    # Data files will be saved automatically.
    timer = tb.Timer()
    timer.tic("corr_dos")
    corr_dos = solver.calc_corr_dos()
    timer.toc("corr_dos")

    # Report time usage and visualization
    if solver.is_master:
        timer.report_total_time()
        analyzer = tb.Analyzer(f"{solver.config.prefix}_info.dat")
        eng, dos = analyzer.calc_dos(corr_dos)
        vis = tb.Visualizer()
        vis.plot_dos(eng, dos)
\end{lstlisting}
We also need to build the model, create a solver, and configure the calculation parameters. Afterwards, we call the \api{calc\_corr\_dos} method to calculate the correlation function and analyze it to obtain DOS. The differences with respect to version 1.3 are that the \api{Solver} class has been renamed to \api{TBPMSolver} for clarity, and \api{config} is now a built-in attribute of solver object. In other words, there is no need to instantiate a \api{config} object as is required in version 1.3. Also, the \api{Analyzer} no longer relies on the model and config but extracts necessary parameters from the data file generated during the calculation. The C++ version of function is defined as
\begin{lstlisting}[language=c++]
void test_tbpm_dos()
{
    // Build the model
    model_t model = make_graphene_sample({ 512, 512, 1 });

    // Create a solver for the model
    TBPMSolver<model_t> solver(model);

    // Set up parameters of the solver
    solver.config.prefix = "graphene";
    solver.config.num_random_samples = 1;
    solver.config.rescale = 9.0;
    solver.config.num_time_steps = 1024;

    // Call 'calc_corr_dos' method of solver to evaluate DOS correlation function
    // Data files will be saved automatically.
    Timer timer;
    timer.tic("dos");
    solver.calc_corr_dos();
    timer.toc("dos");

    // Report time usage
    if (solver.is_master()) {
        timer.report_total_time();
    }
}
\end{lstlisting}
Similar as plotting the band structure, the data files can also be loaded by the I/O functions
\begin{lstlisting}[language=python]
def plot_dos(prefix: str) -> None:
    corr = tb.load_corr_dos(prefix)
    analyzer = tb.Analyzer(f"{prefix}_info.dat")
    energies, dos = analyzer.calc_dos(corr)
    vis = tb.Visualizer()
    vis.plot_dos(energies, dos)
\end{lstlisting}
More examples can be found in the \api{samples/speedtest/plot\_tbpm.py} script of \api{tbplas-cpp}.

\subsubsection{Running the examples}
To run the Python example program, change to \api{samples/demo} directory of \api{tbplas-cpp} and invoke \api{demo.py}
\begin{lstlisting}[language=bash]
cd tbplas-cpp-VERSION_CPP/samples/demo
./demo.py
\end{lstlisting}
For the C++ version, firstly rebuild \api{tbplas-cpp}, then change to the \api{bin} directory and invoke \api{demo}
\begin{lstlisting}[language=bash]
cd tbplas-cpp-VERSION_CPP/build/bin
./demo
\end{lstlisting}
The example programs will take tens of seconds or a few minutes to finish, depending on the hardware. The Python version will plot the results on-the-fly. For the C++ version, visualize the results by
\begin{lstlisting}[language=bash]
PATH_TO_TBPLAS_CPP/samples/speedtest/plot_diag.py graphene_bands
PATH_TO_TBPLAS_CPP/samples/speedtest/plot_tbpm.py graphene_corr_dos
\end{lstlisting}
with \api{PATH\_TO\_TBPLAS\_CPP} replaced by the actual path of unpacked \api{tbplas-cpp-VERSION\_CPP} source code.

\subsubsection{Performance guidance}
In this section, we discuss the guidelines for choosing the optimal parallelization configuration. Since the parallel implementation of version 2.0 is the same as 1.3, the empirical rules established for version 1.3 should still work. For diagonalization-based algorithms, pure MPI-based parallelization is recommended when sufficient memory is available, as matrix diagonalization cannot be efficiently parallelized across OpenMP threads. In the case of TBPM algorithms, OpenMP and MPI parallelization exhibit comparable scaling performance. Therefore, either pure OpenMP or hybrid MPI+OpenMP parallelization is suitable. The optimal number of processes and threads should be determined through benchmarking. Setting the number of MPI processes to the number of CPU sockets and the number of OpenMP threads to the number of physical cores in each socket is typically a good starting point.

\subsection{Advanced modeling}
\label{advanced_modeling}
In this section, we demonstrate the usage of C++ API for advanced modeling taking bilayer graphene quasicrystal as example. The model has a 12-fold symmetry and is formed by twisting one layer by $\frac{\pi}{6}$ with respect to the center $\mathbf{c}=\frac{2}{3}\mathbf{a}_1+\frac{2}{3}\mathbf{a}_2$, where $\mathbf{a}_1$ and $\mathbf{a}_2$ are the lattice vectors of the primitive cell of fixed layer. We construct the model at both \api{PrimitiveCell} and \api{Sample} levels. For clarity only the essential part of the program is shown, while the complete program is located in \api{model.cpp} and \api{model\_sample.cpp} in the \api{samples/speedtest} directory of \api{tbplas-cpp}.

\subsubsection{PrimitiveCell}
\label{advanced_modeling_prim_cell}
Constructing quasicrystal using the C++ API is the same as Python API \cite{Li2023}. Firstly, we define the lattice constant, interlayer distance, twisting angle and center. The radius is passed as a function argument and needs no definition
\begin{lstlisting}[language=c++]
double a = 0.142;
double shift = 0.3349;
double angle = 30.0 / 180.0 * PI;
Eigen::Vector3d center { { 2.0 / 3 }, { 2.0 / 3 }, { 0.0 } };
\end{lstlisting}
We need a large cell to hold the quasicrystal, whose dimension is defined in \api{dim} and can be estimated as $\frac{r}{0.75a}$
\begin{lstlisting}[language=c++]
int rmin_dia = static_cast<int>(std::ceil(radius / (0.75 * a))) + 1;
std::tuple<int, int, int> dim = { rmin_dia, rmin_dia, 1 };
\end{lstlisting}
After introducing the parameters, we build the fixed and twisted layers by calling \api{make\_graphene\_diamond} and \api{extend\_prim\_cell} in the same approach as Python API. The former function is to build the primitive cell of monolayer graphene and the latter is to extend the cell to desired dimension
\begin{lstlisting}[language=c++]
model_t prim_cell = make_graphene_diamond();
model_t layer_fixed = extend_prim_cell(prim_cell, dim);
model_t layer_twisted = extend_prim_cell(prim_cell, dim);
\end{lstlisting}
Then we remove the orbitals falling out of the quasicrystal radius by calling the \api{cutoff\_pc} function. The source code can be found in \ref{src:cutoff_pc}.
\begin{lstlisting}[language=c++]
// Get the Cartesian coordinate of rotation center
center[0] += static_cast<int>(std::get<0>(dim) / 2);
center[1] += static_cast<int>(std::get<1>(dim) / 2);
center = (center.transpose() * prim_cell.get_lattice()).transpose();

// Remove unnecessary orbitals
cutoff_pc(layer_fixed, center, radius);
cutoff_pc(layer_twisted, center, radius);
\end{lstlisting}
After cutting off the layers, we shift and rotate the twisted layer with respect to the center and reshape it to the lattice vectors of fixed layer, which is done by calling \api{spiral\_prim\_cell} and \api{reset\_lattice}
\begin{lstlisting}[language=c++]
spiral_prim_cell(layer_twisted, angle, center, shift);
layer_twisted.reset_lattice(layer_fixed.get_lattice(),
    layer_fixed.get_origin(), 1.0, true);
\end{lstlisting}
Then we merge the layers by calling \api{merge\_prim\_cell} and extend the hopping terms with a cutoff of 0.75 nm
\begin{lstlisting}[language=c++]
std::vector<const PCInterHopping<complex_t>*> inter_hops = {};
merged_cell = merge_prim_cell(prim_cells, inter_hops);
extend_hop(merged_cell, 0.75);
\end{lstlisting}
The \api{extend\_hop} function adds hopping terms according to to Slater-Koster formulation \cite{PhysRevB.86.125413} and the source code can be found in 
\ref{src:extend_hop}. Finally, we save the model to disk
\begin{lstlisting}[language=c++]
merged_cell.save("quasi_crystal_prim_cell");
\end{lstlisting}
The model can be visualized using the \api{plot\_model.py} script
\begin{lstlisting}[language=bash]
PATH_TO_TBPLAS_CPP/python/plot_model.py quasi_crystal_prim_cell --hop-eng-cutoff=0.3
\end{lstlisting}
The argument \api{{-}{-}hop-eng-cutoff} specifies that only the hopping terms larger than 0.3 eV will be shown. The output should be similar to Fig. \ref{fig:poly_mono}(b).

\subsubsection{Sample}
\label{advanced_modeling_sample}
As discussed in Section \ref{rev_model_cpp}, the \api{SuperCell}, \api{SCHopping} and \api{Sample} classes are designed following the \textit{filter-modifier} pattern. The orbitals are firstly generated by populating the supercell, then filtered by the filters to remove the unwanted orbitals and modified by the modifiers to update the orbital positions and energies. The hopping terms are handled in a similar approach. In the case of quasicrystal, the orbital filter should remove the orbitals falling out of the radius, while the orbital modifier should twist and shift the top layer. A hopping modifier is also needed for setting up the intra- and inter-supercell hopping terms according to Slater-Koster formulation. So we begin with defining the filters and modifiers.

The orbital filter should inherit from the \api{AbstractOrbitalFilter} class and overwrite the \api{act} function. The attributes \api{lattice}, \api{origin}, \api{center} and \api{radius} (underscores omitted) define the geometric parameters of the quasicrystal. In the \api{act} function the orbitals are filtered according to their distances to the geometry center, and only those falling within the radius of the quasicrystal will be reserved
\begin{lstlisting}[language=c++]
class CircleFilter final : public AbstractOrbitalFilter {
public:
    void act(std::vector<Orbital>& full_orbitals) const final
    {
        std::vector<Orbital> new_orbitals;
        new_orbitals.reserve(full_orbitals.size());
        for (const auto& orb : full_orbitals) {
            Eigen::Vector3d cart_pos =
                (orb.get_position().transpose() * lattice_).transpose() + origin_;
            Eigen::Vector3d dr = cart_pos - center_;
            if (dr.norm() <= radius_) {
                new_orbitals.push_back(orb);
            }
        }
        new_orbitals.shrink_to_fit();
        full_orbitals = std::move(new_orbitals);
    }
private:
    Eigen::Matrix3d lattice_ = Eigen::Matrix3d::Identity();
    Eigen::Vector3d origin_ = Eigen::Vector3d::Zero();
    Eigen::Vector3d center_ = Eigen::Vector3d::Zero();
    double radius_ = 0.0;
};
\end{lstlisting}

The orbital modifier should inherit from the \api{AbstractOrbitalModifier} class and overwrite the \api{act} function, with \api{center}, \api{angle}, \api{shift} also being geometric parameters of the quasicrystal. In the \api{act} function the orbital positions are updated \textit{in-place} by calling \api{rotate\_coord} and shifted along $z$-axis by the inter-layer distance
\begin{lstlisting}[language=c++]
class TwistModifier final : public AbstractOrbitalModifier {
public:
    void act(OrbitalData& orb_data) const final
    {
        orb_data.orb_pos = rotate_coord(orb_data.orb_pos, angle_, "z", center_);
        orb_data.orb_pos.colwise() += Eigen::Vector3d(0.0, 0.0, shift_);
    }
private:
    Eigen::Vector3d center_ = Eigen::Vector3d::Zero();
    double angle_ = 0.0;
    double shift_ = 0.0;
};
\end{lstlisting}

The hopping modifier should inherit from the \api{AbstractHoppingModifier} class and overwrite the \api{act} function, which updates the hopping energies according to the Slater-Koster formulation \cite{PhysRevB.86.125413}. The \api{calc\_hop} function is defined in \ref{src:calc_hop}
\begin{lstlisting}[language=c++]
class SKTable final : public AbstractHoppingModifier {
public:
    void act(
        const OrbitalData& data_bra,
        const OrbitalData& data_ket,
        HoppingData& hop_data) const final
    {
        for (size_t i = 0; i < hop_data.get_num_hopping(); ++i) {
            hop_data.hop_eng[i] = calc_hop(hop_data.dr.col(i));
        }
    }
};
\end{lstlisting}

After the definition of the filters and modifiers, we define the geometric parameters, estimate the dimension and calculate the Cartesian coordinate of the center as in Section \ref{advanced_modeling_prim_cell}
\begin{lstlisting}[language=c++]
// Geometric parameters
double a = 0.142;
double shift = 0.3349;
double angle = 30.0 / 180.0 * PI;
Eigen::Vector3d center { { 2.0 / 3 }, { 2.0 / 3 }, { 0.0 } };

// Estimate dim for diamond-shaped prim_cell
int rmin_dia = static_cast<int>(std::ceil(radius / (0.75 * a))) + 1;
dim_t dim = { rmin_dia, rmin_dia, 1 };

// Get the Cartesian coordinate of rotation center
center[0] += static_cast<int>(std::get<0>(dim) / 2);
center[1] += static_cast<int>(std::get<1>(dim) / 2);
center = (center.transpose() * prim_cell->get_lattice()).transpose();
\end{lstlisting}
Then we create the fixed and twisted layers of quasicrystal and assign the filters and modifiers to them. Both layers need the orbital filter to remove unwanted orbitals. For the top (twisted) layer, an orbital modifier is essential to shift and twist it with respect to the bottom (fixed) layer
\begin{lstlisting}[language=c++]
// Make layers
using sc_t = builder::SuperCell<complex_t>;
pbc_t pbc = { false, false, false };
auto prim_cell = std::make_shared<model_t>(make_graphene_diamond());
auto sc_fixed = std::make_shared<sc_t>(prim_cell, dim, pbc);
auto sc_twisted = std::make_shared<sc_t>(prim_cell, dim, pbc);

// Make filters and modifiers and assign to the layers
auto circle = std::make_shared<CircleFilter>(
    sc_fixed->get_sc_lattice(), sc_fixed->get_sc_origin(), center, radius);
auto twist = std::make_shared<TwistModifier>(center, angle, 0.3349);
sc_fixed->add_filter(circle);
sc_twisted->add_filter(circle);
sc_twisted->add_modifier(twist);
\end{lstlisting}
Then we create the intra-supercell hopping containers of each layer, and the inter-supercell hopping container between the layers. Both containers should be instantiated from the \api{SCHopping} class, and differ only in the arguments of the constructor. Since intra- and inter-supercell hopping containers are treated on the same footing, they all need the Slater-Koster hopping modifier
\begin{lstlisting}[language=c++]
// Make intra and inter hopping generators
using sc_hop_t = builder::SCHopping<complex_t>;
auto hop_fixed = std::make_shared<sc_hop_t>(sc_fixed, sc_fixed, 0.75);
auto hop_twisted = std::make_shared<sc_hop_t>(sc_twisted, sc_twisted, 0.75);
auto hop_inter = std::make_shared<sc_hop_t>(sc_fixed, sc_twisted, 0.75);

// Make and assign modifiers
auto sk = std::make_shared<SKTable>();
hop_fixed->add_modifier(sk);
hop_twisted->add_modifier(sk);
hop_inter->add_modifier(sk);
\end{lstlisting}
Finally, we assemble the layers and containers into a sample and save it to disk
\begin{lstlisting}[language=c++]
Sample<complex_t> sample({ sc_fixed, sc_twisted },
    { hop_fixed, hop_twisted, hop_inter });
sample.init_array();
sample.save("quasi_crystal_sample");
\end{lstlisting}
The model can be visualized in the same approach as primitive cell
\begin{lstlisting}[language=bash]
PATH_TO_TBPLAS_CPP/python/plot_model.py quasi_crystal_sample --hop-eng-cutoff=0.3
\end{lstlisting}

\subsection{New features}
\subsubsection{Spin texture}
In this section, we demonstrate the usage of \api{SpinTexture} class to calculate the spin texture of Kane-Mele model \cite{kane_mele}. This is done in the \api{test\_spin} function in \api{samples/speedtest/diag.py} of \api{tbplas-cpp}, which is defined as
\begin{lstlisting}[language=python]
def test_spin():
    # Import the model from repository and rotate the model by pi/6
    model_graph = tb.make_graphene_soc()
    model_graph.rotate(np.pi / 6)

    # Create solver and set params
    solver_graph = tb.SpinTexture(model_graph)
    solver_graph.config.prefix = "kane_mele"
    solver_graph.config.k_points = 2 * (tb.gen_kmesh((36, 36, 1)) - 0.5)
    solver_graph.config.k_points[:, 2] = 0.0
    solver_graph.config.spin_major = False

    # Calculation
    data = solver_graph.calc_spin_texture()

    # Plot
    if solver_graph.is_master:
        vis = tb.Visualizer()
        plot_sigma_band(data, vis)
        plot_sigma_eng(data, vis)
\end{lstlisting}
Firstly, we import the model from repository with the \api{make\_graphene\_soc} and rotate it by $\frac{\pi}{6}$ counter-clockwise for better appearance of the Brillouin zone. Then we create a solver from \api{SpinTexture} class and set up the parameters. The k-points are sampled with $k_a,k_b\in[-1, 1]$ and $k_c=0$ with dimension of $36\times36\times1$. The parameter \api{spin\_major} controls whether the orbitals are arranged in spin-major order, i.e. $(\phi_{1\uparrow}, \phi_{2\uparrow}, ..., \phi_{n\uparrow}, \phi_{1\downarrow}, \phi_{2\downarrow}, ..., \phi_{n\downarrow})$. Otherwise, the orbital order will be $(\phi_{1\uparrow}, \phi_{1\downarrow}, \phi_{2\uparrow}, \phi_{2\downarrow}, ..., \phi_{n\uparrow}, \phi_{n\downarrow})$. Finally we call the \api{calc\_spin\_texture} method to get the expectation values of Pauli operators and visualize them using the \api{Visualizer} class.

The function \api{plot\_sigma\_band} plots the spin texture of specific band, while \api{plot\_sigma\_eng} plot the spin texture of specific energy range. Take \api{plot\_sigma\_band} for example. In this function we firstly extract $\langle\sigma_x\rangle$, $\langle\sigma_y\rangle$ and $\langle\sigma_z\rangle$ for given band, then plot $\langle\sigma_z\rangle$ as scalar field of $\mathbf{k}$-point using the \api{plot\_scalar} method of \api{Visualizer} class and $(\langle\sigma_x\rangle, \langle\sigma_y\rangle)$ as vector field using \api{plot\_vector}
\begin{lstlisting}[language=python]
def plot_sigma_band(
    spin_data: tb.SpinData,
    vis: tb.Visualizer,
    ib: int = 0) -> None:
    kpt_cart = spin_data.kpt_cart
    sigma_x = spin_data.sigma_x[:, ib]
    sigma_y = spin_data.sigma_y[:, ib]
    sigma_z = spin_data.sigma_z[:, ib]
    vis.plot_scalar(x=kpt_cart[:, 0], y=kpt_cart[:, 1], z=sigma_z, scatter=True,
                    num_grid=(480, 480), cmap="jet", with_colorbar=True)
    vis.plot_vector(x=kpt_cart[:, 0], y=kpt_cart[:, 1], u=sigma_x, v=sigma_y,
                    cmap="jet", with_colorbar=False)
\end{lstlisting}

\begin{figure}[h]
	\centering
	\includegraphics[width=1.0\linewidth]{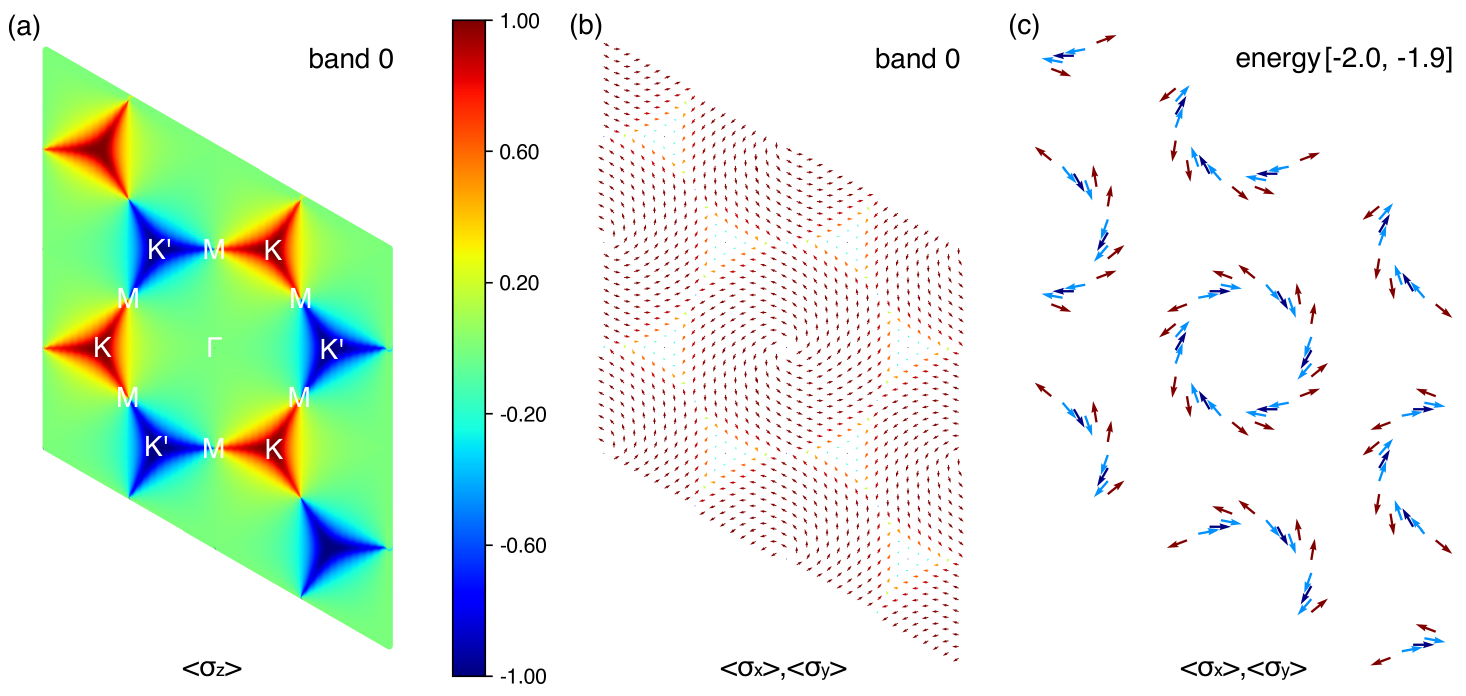}
	\caption{Spin texture of Kane-Mele model. (a) Expectation value of $\sigma_z$ of the first band as scalar field evaluated on a $640\times640\times1$ $\mathbf{k}$-grid. (b) Expectation value of $\sigma_x$ and $\sigma_y$ of the first band as vector field evaluated on a $36\times36\times1$ $\mathbf{k}$-grid. (c) is similar to (b), but for states within the energy range of [-2.0, -1.9]. K, M and $\Gamma$ denote the high symmetric $\mathbf{k}$-points in the first Brillouin zone.}
	\label{fig:spin_texture}
\end{figure}

The example can be run as
\begin{lstlisting}[language=bash]
PATH_TO_TBPLAS_CPP/samples/speedtest/speedtest.py spin
\end{lstlisting}
The output is shown in Fig. \ref{fig:spin_texture}, where non-trivial textures due to spin-orbital coupling can be observed. The expectation value of $\sigma_z$ reaches its extrema with opposite signs at $\mathbf{K}$ and $\mathbf{K^\prime}$ points of the Brillouin zone, while decreasing to zero at $\mathbf{M}$ point. The clockwise (blue) and counter-clockwise (brown) spin orientations around $\mathbf{\Gamma}$ point in Fig.\ref{fig:spin_texture}(c) clearly show the effects of Rashba spin-orbital coupling.

The C++ version of the example is located in \api{samples/speedtest/diag.cpp} of \api{tbplas-cpp}, which is much similar to the Python version and will not be shown for clarity. The program can be invoked by
\begin{lstlisting}[language=bash]
PATH_TO_TBPLAS_CPP/samples/speedtest/speedtest spin
\end{lstlisting}
And the results can be plotted by
\begin{lstlisting}[language=bash]
PATH_TO_TBPLAS_CPP/samples/speedtest/plot_diag.py kane_mele_spin
\end{lstlisting}
The output is consistent with the Python version.

\subsubsection{Berry curvature and Chern number}
\label{usage_berry}
In this section, we demonstrate the usage of \api{Berry} class to calculate the Berry curvature and Chern number of Haldane model \cite{haldane, coh_python_2022}. The calculation is done in the \api{test\_berry} function in \api{samples/speedtest/diag.py} of \api{tbplas-cpp}, which is defined as
\begin{lstlisting}[language=python]
def test_berry():
    # Build the model and output analytical Hamiltonian
    model = make_haldane()
    model.print_hk(convention=1, output_format="cpp")

    # Create solver and set params
    solver = tb.Berry(model)
    solver.config.k_grid_size = (120, 120, 1)
    solver.config.bz_size = (2, 2, 1)
    solver.config.bz_shift = np.array([-1.0, -1.0, 0.0])
    solver.config.num_occ = 1
    solver.config.ham_deriv_analytical = True

    # Calculate Berry curvature using Kubo formula
    solver.config.prefix = "haldane_kubo"
    data_kubo = solver.calc_berry_curvature_kubo()

    # Calculate Berry curvature using Wilson loop
    solver.config.prefix = "haldane_wilson"
    data_wilson = solver.calc_berry_curvature_wilson()

    # Plot
    if solver.is_master:
        vis = tb.Visualizer()
        plot_omega_xy(data_kubo, vis)
        plot_omega_xy(data_wilson, vis)
\end{lstlisting}
We firstly build the model with the \api{make\_handane} function defined in \api{model.py} and get the analytical Hamiltonian for later use in Section \ref{analy_ham}. Then we create the solver from \api{Berry} class and set the parameters. Similar to spin texture, we also need to sample the Brillouin zone with a $\mathbf{k}$-grid. But we cannot use $k_a,k_b\in[-1, 1]$ directly as for spin texture since the Chern number is sensitive to Brillouin zone size. Instead, we set the size of Brillouin zone with the \api{bz\_size} argument and shift the $\mathbf{k}$-points by a vector of $\mathbf{b}=-\mathbf{b}_1-\mathbf{b}_2$ with $\mathbf{b}_1$ and $\mathbf{b}_2$ denoting the basis vectors of reciprocal lattice. The final $\mathbf{k}$-points for Berry curvature calculation is thus $k_a,k_b\in[-1, 1]$ and $k_c=0$ with dimension of $240\times240\times1$. The argument \api{num\_occ} defines the size of $M_{nm}^{\mathbf{k}_i,\mathbf{k}_{i+1}}$ defined in Eqn. \ref{eq:unk_ovelap}. If it is set to 1, the Berry curvature for the first band will be produced. Otherwise, we will get the total Berry curvature for all occupied bands. The argument \api{ham\_deriv\_analytical} defines whether to use analytical derivation of the Hamiltonian with respect to $\mathbf{k}$-point or numerical derivation.

\begin{figure}[h]
	\centering
	\includegraphics[width=1.0\linewidth]{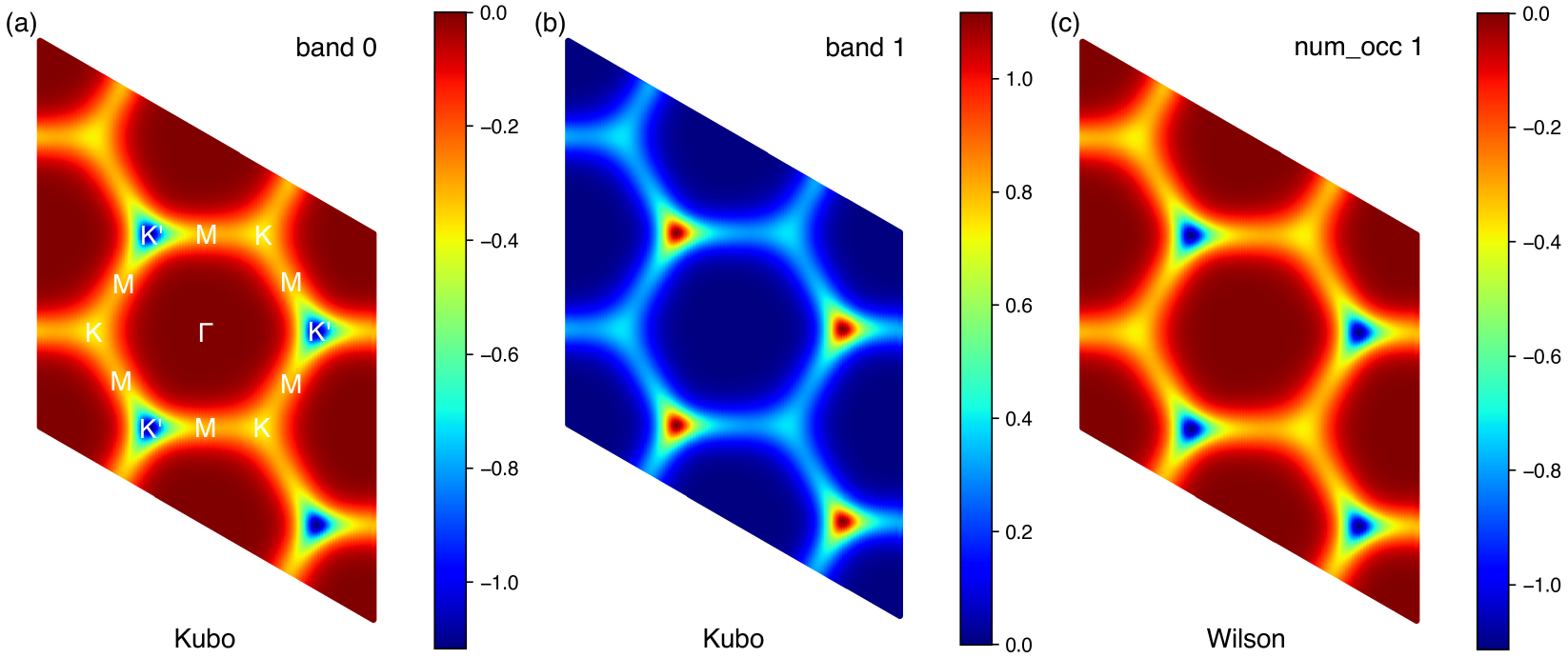}
	\caption{Berry curvature of Haldane model. Results for (a) band 0 and (b) band 1 are obtained using Kubo formula, while (c) is obtained using Wilson loop with num\_occ set to 1. The calculations are performed on a $240\times240\times1$ $\mathbf{k}$-grid and the unit of Berry curvature is $\mathrm{nm}^2$. K, M and $\Gamma$ denote the high symmetric $\mathbf{k}$-points in the first Brillouin zone.}
	\label{fig:berry}
\end{figure}

After setting the parameters, we calculate the Berry curvature using the Kubo formula and Wilson loop method by calling the \api{calc\_berry\_curvature\_kubo} and \api{calc\_berry\_curvature\_wilson} methods of \api{Berry} class, respectively. Note that we use different output prefixes to avoid overwriting the data files. Finally, we utilize the \api{Visualizer} class to plot the Berry curvature. The function \api{plot\_omega\_xy} is similar to the function for plotting spin texture and the source code will not be shown for clarity. The example can be invoked as
\begin{lstlisting}[language=bash]
PATH_TO_TBPLAS_CPP/samples/speedtest/speedtest.py berry
\end{lstlisting}
The results are shown in Fig. \ref{fig:berry}. It is clear that the Berry curvature of Haldane model gets its extrema at either $\mathbf{K}$ or $\mathbf{K}^\prime$ points depending on the band index, and the Wilson loop method produces exactly the same result as Kubo formula for the first band if only one band is taken into consideration. The Chern numbers will be print to stdout during the calculation. We can observe that the first and second bands have different Chern numbers due to the opposite signs of Berry curvatures, and the Wilson loop method predicts the same Chern number as Kubo formula for the first band (irrelevant messages omitted)
\begin{lstlisting}
Chern number for band 0: -1
Chern number for band 1: 1
Chern number for num_occ 1: -1
\end{lstlisting}

The C++ version of the example is located in \api{samples/speedtest/diag.cpp} of \api{tbplas-cpp}, which will not be shown for clarity. The program can be invoked by
\begin{lstlisting}[language=bash]
PATH_TO_TBPLAS_CPP/samples/speedtest/speedtest berry
\end{lstlisting}
And the results can be plotted by
\begin{lstlisting}[language=bash]
PATH_TO_TBPLAS_CPP/samples/speedtest/plot_diag.py haldane_kubo_berry
PATH_TO_TBPLAS_CPP/samples/speedtest/plot_diag.py haldane_wilson_berry
\end{lstlisting}
The output is consistent with the Python version.

\subsubsection{Analytical Hamiltonian}
\label{analy_ham}
In this section, we reproduce the Berry curvature and Chern numbers of Haldane model using the analytical Hamiltonian from Section \ref{usage_berry}. We achieve this by defining a model class \api{HaldaneHK} as derived class of \api{AnalyticalModel} and overwrite the \api{build\_ham\_dense} method. The source code can be found in \api{model.h} and \api{model.cpp} in \api{samples/speedtest} of \api{tbplas-cpp}, and will not be shown here for clarity. We focus on the usage of the model class as demonstrated in the \api{test\_berry} function of \api{diag.cpp}
\begin{lstlisting}[language=c++]
void test_berry()
{
    HaldaneHK model;
    Berry<HaldaneHK> solver(model);
    solver.config.k_grid_size = { 120, 120, 1 };
    solver.config.bz_size = {2, 2, 1};
    solver.config.bz_shift = Eigen::Vector3d(-1.0, -1.0, 0.0);
    solver.config.num_occ = 1;
    solver.config.ham_deriv_analytical = false;
    solver.config.prefix = "haldane_kubo";
    auto data_kubo = solver.calc_berry_curvature_kubo();
    solver.config.prefix = "haldane_wilson";
    auto data_wilson = solver.calc_berry_curvature_wilson();
}
\end{lstlisting}
Firstly, the \api{HaldaneHK} class is instantiated to yield a Haldane model. Then a \api{Berry} solver is created taking the \api{HaldaneHK} class as template argument. The other parts are much similar to the Python program in Section \ref{usage_berry}. The only difference is that the \api{ham\_deriv\_analytical} argument should be set to false, since we have not overwritten the \api{build\_ham\_der\_dense} method to evaluate the analytical Hamiltonian derivatives. The invocation of the program, the data plotting procedure and the results are the same to the C++ program in Section \ref{usage_berry}, and the results are consistent with the Python version. \section{Benchmarks}
\label{benchmark}

\def\speedup#1{\textbf{#1$\boldsymbol{\times}$}}

\def\tabfmt#1{#1}

\subsection{Modeling tools}
\label{benchmark_model}
In this section, we benchmark the modeling tools of \api{TBPLaS} 2.0 against version 1.3. Both the Python/Cython and C++ modeling tools are tested at \api{PrimitiveCell} and \api{Sample} levels. We consider three kinds of models: twisted bilayer graphene (TBG), twisted bilayer graphene quasicrystal and Si\'{e}rpinski carpet fractal. The details and the algorithms for constructing the models can be found in the article for \api{TBPLaS} 1.3 \cite{Li2023}. The sizes of TBG, quasicrystal and fractal are controlled by the twisting index $i$, the radius $r$ and the iteration number $n$, respectively. For TBG and quasicrystal, the number of orbitals and hopping terms scale as $i^2$ and $r^2$, while for fractal they scale as $L^{2n}$ with $L$ being the dimension of iteration pattern. The sizes of models employed for the benchmarks are summarized in Table \ref{tab:model_size}. We consider TBG with twisting index ranging from 20 to 100, leading to model sizes of 5k-121k. The quasicrystals have radius of 6-30 nm and 8k-215k orbitals. For the Python/Cython modeling tools we consider fractals with iteration number $n\le5$, since the time usage of larger models will be unaffordable. For the C++ tools, we further increase the iteration number to 7, leading to a model with 8 million orbitals and 29 million hopping terms.

\begin{table}[htbp]
	\begin{center}
		\caption{Summary of the numbers of orbitals and hopping terms of the models employed in the benchmarks. The term \textit{parameter} indicates the twisting index $i$ of TBG, the radius $r$ of quasicrystal and the iteration number $n$ of fractal depending on the model type. Quasicrystal radius $r$ is in nanometer.}
		\label{tab:model_size}
		\begin{tabular}{ccrr}
\hline\hline
            Model & Parameter & \makecell{Number of \\ orbitals} & \makecell{Number of \\hopping terms} \\
			\hline
\multirow{5}{*}{TBG}& 20 & 5,044 & 302,635 \\
                             & 40 & 19,684 & 1,180,894 \\
                             & 60 & 43,924 & 2,635,255 \\
                             & 80 & 77,764 & 4,665,430 \\
                             & 100 & 121,204 & 7,271,587 \\
            \hline
\multirow{5}{*}{quasicrystal} & 6 & 8,592 & 489,168 \\
                                        & 12 & 34,392 & 2,010,684 \\
                                        & 18 & 77,496 & 4,570,152 \\
                                        & 24 & 137,808 & 8,162,112 \\
                                        & 30 & 215,556 & 12,800,268 \\
            \hline
\multirow{5}{*}{fractal} & 3 & 2,048 & 6,852 \\
                                  & 4 & 16,384 & 56,100 \\
                                  & 5 & 131,072 & 452,676 \\
                                  & 6 & 1,048,576 & 3,633,060 \\
                                  & 7 & 8,388,608 & 29,099,460 \\
			\hline\hline
		\end{tabular}
	\end{center}
\end{table}

The time usage and speedup of modeling tools are summarized in Table \ref{tab:model_time} and \ref{tab:model_speedup}. As aforementioned in Section \ref{intro}, the Cython-based modeling tools of version 1.3 are inefficient for monolithic models. This can be proved by the time usage of 1.3 Python and 1.3 Cython in Table \ref{tab:model_time}, where the latter is 2-3 times larger than the former for monolithic TBG and quasicrystal, but an order of magnitude lower for polylithic fractal models. This inefficiency has been fixed in version 2.0, with Cython tools much faster than the Python tools even for monolithic models. Comparing the same tool of version 2.0 to 1.3, the Python \api{PrimitiveCell} class has speedup of \speedup{1.364}-\speedup{1.619} (36.4\%-61.9\%) depending on the model type. The Cython \api{Sample} class has speedup of \speedup{2.269}-\speedup{9.375} for monolithic models and \speedup{22.378}-\speedup{385.635} for polylithic models (fractal with $n=3$  neglected due to the insufficiently accurate time measurements). All speedup indicates \todo{significant improvements} of the existing Python/Cython modeling tools.

\begin{table}[htbp]
  \centering
  \caption{Time usage of modeling at \api{PrimitiveCell} and \api{Sample} levels using different APIs for \api{TBPLaS} 1.3 and 2.0. The convention for term \textit{parameter} follows Table. \ref{tab:model_size}. The programs have been compiled with GCC 11.4.0 at \api{-O3} level and performed on a computer with 2 Intel Xeon Gold 6548Y+ processors and 256GB RAM. Some non-C++ tests for fractal have been skipped due to the unaffordable time usage.}
  \label{tab:model_time}
\begin{tabular}{ccrrrrrr}
    \hline\hline
    \multirow{2}{*}{\tabfmt{Model}} & 
    \multirow{2}{*}{\tabfmt{Parameter}} & 
    \multicolumn{3}{c}{\tabfmt{PrimitiveCell (s)}} &
    \multicolumn{3}{c}{\tabfmt{Sample (s)}} \\
    \cmidrule(lr){3-5} \cmidrule(lr){6-8}
    & & \tabfmt{1.3 Python} & \tabfmt{2.0 Python} & \tabfmt{2.0 C++} & \tabfmt{1.3 Cython} & \tabfmt{2.0 Cython} & \tabfmt{2.0 C++} \\
    \hline
    \multirow{5}{*}{\tabfmt{TBG}} 
    & 20 & 6.869 & 4.608 & 0.364 & 7.839 & 3.455 & 0.168 \\
    & 40 & 27.463 & 19.251 & 1.828 & 38.770 & 14.671 & 0.771 \\
    & 60 & 61.973 & 43.631 & 4.745 & 112.072 & 34.696 & 2.197 \\
    & 80 & 111.799 & 79.446 & 9.553 & 257.463 & 63.617 & 4.334 \\
    & 100 & 175.307 & 126.864 & 15.987 & 520.900 & 101.544 & 7.186 \\
    \hline
    \multirow{5}{*}{\tabfmt{quasicrystal}} 
    & 6 & 9.526 & 6.887 & 0.624 & 13.113 & 5.065 & 0.277 \\
    & 12 & 47.266 & 34.393 & 3.278 & 77.524 & 23.032 & 1.596 \\
    & 18 & 133.049 & 93.544 & 9.013 & 258.174 & 55.527 & 4.153 \\
    & 24 & 295.420 & 199.877 & 17.551 & 667.020 & 101.702 & 8.414 \\
    & 30 & 600.496 & 370.858 & 30.801 & 1535.989 & 163.837 & 14.065 \\
    \hline
    \multirow{5}{*}{\tabfmt{fractal}} 
    & 3 & 0.146 & 0.107 & 0.007 & 0.011 & 0.022 & 0.003 \\
    & 4 & 8.140 & 5.868 & 0.037 & 0.828 & 0.037 & 0.016 \\
    & 5 & 705.456 & 513.316 & 0.272 & 81.369 & 0.211 & 0.075 \\
    & 6 & - & - & 3.033 & - & - & 0.747 \\
    & 7 & - & - & 34.815 & - & - & 6.694 \\
    \hline\hline
  \end{tabular}
\end{table}

\begin{table}[htbp]
    \centering
    \caption{Speedup of the modeling tools. The convention for term \textit{parameter} follows Table. \ref{tab:model_size}. Columns 3-4 are the speedup of Python/Cython APIs of version 2.0 versus version 1.3. Columns 5-6 are the speedup of C++ APIs versus Python counterparts for version 2.0. Column 7 is the speedup of C++ \api{PrimitiveCell} versus {Sample} for version 2.0. The speedup is defined as the inverse ratio of time usage, i.e., $A/B:=t_B/t_A$}.
    \label{tab:model_speedup}

\begin{tabular}{cccrrrc}
      \hline\hline
      \multirow{2}{*}{\tabfmt{Model}} & 
      \multirow{2}{*}{\tabfmt{Parameter}} & 
      \multicolumn{2}{c}{\tabfmt{2.0 / 1.3}} &
      \multicolumn{2}{c}{\tabfmt{2.0 C++ / Python}} &
      \multirow{2}{*}{\makecell{2.0 C++ Sample / \\ PrimitiveCell}} \\
      \cmidrule(lr){3-4} \cmidrule(lr){5-6}
      & & \tabfmt{PrimitiveCell} & \tabfmt{Sample} & \tabfmt{PrimitiveCell} & \tabfmt{Sample} &  \\
      \hline
      \multirow{5}{*}{\tabfmt{TBG}} 
      & 20 & 1.491 & 2.269 & 12.669 & 20.565 & 2.167 \\
      & 40 & 1.427 & 2.643 & 10.531 & 19.029 & 2.371 \\
      & 60 & 1.420 & 3.230 & 9.195 & 15.792 & 2.160 \\
      & 80 & 1.407 & 4.047 & 8.316 & 14.679 & 2.204 \\
      & 100 & 1.382 & 5.130 & 7.935 & 14.131 & 2.225 \\
      \hline
      \multirow{5}{*}{\makecell{\tabfmt{quasicrystal}}} 
      & 6 & 1.383 & 2.589 & 11.037 & 18.285 & 2.253 \\
      & 12 & 1.374 & 3.366 & 10.492 & 14.431 & 2.054 \\
      & 18 & 1.422 & 4.650 & 10.379 & 13.370 & 2.170 \\
      & 24 & 1.478 & 6.559 & 11.388 & 12.087 & 2.086 \\
      & 30 & 1.619 & 9.375 & 12.040 & 11.649 & 2.190 \\
      \hline
      \multirow{5}{*}{\makecell{\tabfmt{fractal}}} 
      & 3 & 1.364 & 0.500 & 15.286 & 7.333 & 2.333 \\
      & 4 & 1.387 & 22.378 & 158.595 & 2.313 & 2.313 \\
      & 5 & 1.374 & 385.635 & 1887.191 & 2.813 & 3.627 \\
      & 6 & - & - & - & - & 4.060 \\
      & 7 & - & - & - & - & 5.201 \\
      \hline\hline
    \end{tabular}\end{table}

The brand new C++ implementation of modeling tools in version 2.0 is \todo{orders of magnitude} faster than the Python/Cython counterparts. The \api{PrimitiveCell} class has speedup of \speedup{7.935}-\speedup{12.659} for monolithic models and \speedup{15.286}-\speedup{1887.378} for polylithic models. For the \api{Sample} class, the speedup is \speedup{11.649}-\speedup{20.565} for monolithic models and \speedup{2.313}-\speedup{2.813} for polylithic models (fractal with $n=3$ neglected due to inaccuracy). The relatively low speedup of \api{Sample} is because the Cython version shares much source code with the C++ version and is already fast enough. The \api{Sample} class is at least twice as fast as the \api{PrimitiveCell} class, achieving the best efficiency among all the modeling tools.

Finally, we suggest the thumb rule for choosing the appropriate modeling tool among all the variants. Python/Cython versions are recommended for users not familiar with C++, or if the modeling efficiency is not a concern. Advanced users are recommended to use the C++ modeling tools, with \api{PrimitiveCell} being adequate in most cases. If extreme efficiency is desired, the C++ \api{Sample} class is the only option.

\subsection {Solvers}
\label{benchmark_solver}
\subsubsection{Diagonalization-based solvers}

In this section, we benchmark the diagonalization-based solvers of versions 2.0 and 1.3 using the Python API. The \api{DiagSolver} class for DOS calculation, \api{Z2} class for $\mathbb{Z}_2$ topological invariant, and \api{Lindhard} class for response functions are tested. We consider the conventional cell of bulk silicon with 32 orbitals per cell \cite{silicon} as the model for calculating DOS and response functions, and the bilayer bismuth with 12 orbitals per cell \cite{bismuth_bilayer} for $\mathbb{Z}_2$ topological invariant. The dimension of $\mathbf{k}$-grid is $32\times32\times32$ for DOS and response functions, and  $2000\times2000\times1$ for $\mathbb{Z}_2$.

The time usage and speedup are summarized in Table \ref{tab:diag_time} and \ref{tab:diag_speedup}. The speedup is in the range of \speedup{1.370}-\speedup{5.883} depending on the calculation type (function), the compiler and the underlying math library. In most cases, the solvers of version 2.0 are \speedup{2}-\speedup{4} times faster than those of version 1.3, indicating significant efficiency improvements. The speedup mainly comes from the reduced overhead of function calls between Python and C++ components, since the diagonalization and post-processing are all done in the C++ core in version 2.0. Another advantage of working in C++ is the reduced RAM usage, as there is no need to store the eigenstates of all $\mathbf{k}$-points simultaneously. In fact, the computer will run out of RAM if a denser $\mathbf{k}$-grid than $32\times32\times32$ is employed for the \api{DiagSolver} and \api{Lindhard} classes of version 1.3.

\begin{table}[htbp]
  \centering
  \caption{Time usage of diagonalization-based solvers for \api{TBPLaS} 1.3 and 2.0. Density of states (DOS), dynamic polarizability (DP) and AC conductivity (AC) have been tested on the same hardware as Table. \ref{tab:model_time}. The compilers are GCC 11.4.0 and Intel oneAPI 2025.2.0 with the \api{-O3} optimization flag. The parallelization configuration is 64 MPI processes $\times$ 1 OpenMP thread per process.}
  \label{tab:diag_time}
\begin{tabular}{crrrrrr}
    \hline\hline
    \multirow{2}{*}{\tabfmt{Function}} & 
    \multicolumn{2}{c}{\tabfmt{1.3 (s)}} & 
    \multicolumn{4}{c}{\tabfmt{2.0 (s)}} \\
    \cmidrule(lr){2-3} \cmidrule(lr){4-7}
    & \tabfmt{GCC} & 
    \tabfmt{Intel} & 
    \tabfmt{GCC} & 
    \tabfmt{\makecell{GCC+\\MKL}} & 
    \tabfmt{Intel} & 
    \tabfmt{\makecell{Intel+\\MKL}} \\
    \hline
    \tabfmt{DOS} & 1.076 & 1.243 & 0.513 & 0.462 & 0.213 & 0.278 \\
    \tabfmt{Z2} & 8.122 & 8.000 & 2.044 & 2.149 & 1.595 & 1.937 \\
    \tabfmt{DP} & 11.952 & 11.927 & 8.240 & 8.271 & 4.874 & 4.850 \\
    \tabfmt{AC} & 7.059 & 6.876 & 5.151 & 4.586 & 3.557 & 2.982 \\
    \hline\hline
  \end{tabular}\end{table}

\begin{table}[htbp]
  \centering
  \caption{Speedup of diagonalization-based solvers of version 2.0 versus version 1.3. The speedup is defined as $t_{1.3}/t_{2.0}$ with the subscripts denoting the versions. For both GCC and GCC+MKL, the time usage of GCC of version 1.3 is taken as the reference. Similar rule holds for Intel and Intel+MKL.}
  \label{tab:diag_speedup}
\begin{tabular}{c*{4}{c}}
    \hline\hline
    \tabfmt{Function} & 
    \tabfmt{GCC} & 
    \tabfmt{Intel} & 
    \tabfmt{\makecell{GCC+\\MKL\\}} & 
    \tabfmt{\makecell{Intel+\\MKL}} \\
    \hline
    \tabfmt{DOS} & 2.096 & 5.833 & 2.332 & 4.463 \\
    \tabfmt{Z2} & 3.974 & 5.017 & 3.779 & 4.129 \\
    \tabfmt{DP} & 1.450 & 2.447 & 1.445 & 2.459 \\
    \tabfmt{AC} & 1.370 & 1.933 & 1.539 & 2.306 \\
    \hline\hline
  \end{tabular}\end{table}

\subsubsection{TBPM Solver}

In this section, we benchmark the \api{TBPMSolver} class of versions 2.0 and 1.3 using the Python API. All the capabilities (functions) of the solver, including LDOS, DOS, dynamic polarizability (DP), AC conductivity (AC), DC conductivity (DC), Hall conductivity (Hall), quasi-eigenstates (QE) and time-dependent wave function (WFT) are tested using a monolayer graphene supercell as the model. The supercell dimension is $1024\times1024\times1$ for DC and Hall conductivity, and $4096\times4096\times1$ for other capabilities. The reason is that DC and Hall conductivity are memory-demanding and the GPU device will run out of VRAM if a larger supercell is employed, making GPU tests impractical.

The time usage and speedup are summarized in Table \ref{tab:tbpm_time} and  \ref{tab:tbpm_speedup}. The speedup is strongly dependent on the computational device, the compiler/math library, and the calculation type (function). Considering the CPU tests, DC and Hall conductivity have the largest speedup of more than \speedup{25} without MKL. If MKL is enabled, DC still has a speedup as large as \speedup{23.134}. LDOS using Haydock recursive method has the third largest speedup of more than \speedup{10} without MKL and \speedup{5.111} with MKL. Other capabilities have speedup of \speedup{2}-\speedup{4} without MKL, and at least \speedup{1.467} with MKL. The reason for the relatively lower speedup with MKL is that MKL is already highly optimized. Similar phenomenon can also be observed in the speedup of GCC and Intel, with the latter lower than the former. In summary, the \api{TBPMSolver} of version 2.0 is \todo{several times or even an order of magnitude faster} than version 1.3 on CPU. The speedup mainly comes from the composite functions that avoid the use of temporary arrays and unnecessary copy assignments. DC and Hall conductivity have further optimizations reusing intermediate results. The different speedup of each calculation type is due to the different number of function calls to the composite functions, and the overall speedup is actually the weighted average of the speedup of all the optimizations.

\begin{table}[htbp]
    \centering
    \caption{Time usage of TBPM solver for \api{TBPLaS} 1.3 and 2.0. Local density of states (LDOS), DOS, DP, AC, quasi-eigenstates (QE) and time-dependent wave function (WFT) have been tested using monolayer graphene supercell as the model. The CPUs tests have been performed on the same hardware as Table \ref{tab:model_time}, and the GPU tests have been done on an NVIDIA A800 graphics card. The compilers and optimization flags are the same to Table \ref{tab:diag_time}. The parallelization configuration is 1 MPI process $\times$ 64 OpenMP threads per process.}
    \label{tab:tbpm_time}
\begin{tabular}{crrrrrrrrr}
       \hline\hline
       \multirow{2}{*}{\tabfmt{Function}} & 
       \multicolumn{3}{c}{\tabfmt{1.3 (s)}} & 
       \multicolumn{5}{c}{\tabfmt{2.0 (s)}} \\
       \cmidrule(lr){2-4} \cmidrule(lr){5-9}
       & \tabfmt{GCC} & \tabfmt{Intel} & \tabfmt{\makecell{Intel+\\MKL}} & \tabfmt{GCC} & \tabfmt{\makecell{GCC+\\MKL}} & \tabfmt{Intel} & \tabfmt{\makecell{Intel+\\MKL}} & \tabfmt{GPU} \\
       \hline
        \tabfmt{LDOS} & 1120.864 & 960.021 & 470.710 & 88.458 & 95.917 & 94.243 & 92.098 & 17.873 \\
        \tabfmt{DOS} & 1989.160 & 1378.443 & 925.724 & 552.503 & 585.698 & 540.672 & 549.818 & 90.816 \\
        \tabfmt{DP} & 7380.289 & 5116.578 & 3456.689 & 2023.009 & 2156.289 & 2064.568 & 2194.147 & 1241.574 \\
        \tabfmt{AC} & 6080.831 & 4330.780 & 2676.322 & 1578.804 & 1814.701 & 1627.581 & 1823.801 & 957.035 \\
        \tabfmt{DC} & 18445.589 & 22286.926 & 10087.947 & 526.052 & 430.043 & 469.914 & 436.060 & 437.786 \\
        \tabfmt{Hall} & 18904.279 & 21057.740 & 689.693 & 750.566 & 594.802 & 665.074 & 604.564 & 1123.126 \\
        \tabfmt{QE} & 4097.116 & 2701.467 & 1995.554 & 1093.656 & 1165.026 & 1131.342 & 1150.582 & 178.475 \\
        \tabfmt{WFT} & 1858.815 & 1301.351 & 910.374 & 539.479 & 601.445 & 558.475 & 548.937 & 87.241 \\
      \hline\hline
    \end{tabular}\end{table}

\begin{table}[htbp]
    \centering
    \caption{Speedup of TBPM solver of version 2.0 versus version 1.3. The speedup is defined as $t_{1.3}/t_{2.0}$ with the subscripts denoting the versions. For GCC+MKL the reference is the time usage of GCC of 1.3. The speedup of GPU is defined as $t_{2.0}^{\mathrm{GCC}}/t_{2.0}^{\mathrm{GPU}}$, and the normalized speedup is further divided by the TFLOPS ratio of GPU to CPU (2.644 for one NVIDIA A800 and two Intel Xeon Gold 6548Y+).}
    \label{tab:tbpm_speedup}
\begin{tabular}{crrrrcc}
      \hline\hline
      \tabfmt{Function} & \tabfmt{GCC} & \tabfmt{\makecell{GCC+\\MKL}} & \tabfmt{Intel} & \tabfmt{\makecell{Intel+\\MKL}} & \tabfmt{GPU / CPU} & \tabfmt{\makecell{GPU / CPU\\normalized}}\\
      \hline
       \tabfmt{LDOS} & 12.671 & 11.686 & 10.187 & 5.111 & 4.949 & 1.872 \\
       \tabfmt{DOS} & 3.600 & 3.396 & 2.550 & 1.684 & 6.084 & 2.301 \\
       \tabfmt{DP} & 3.648 & 3.423 & 2.478 & 1.575 & 1.629 & 0.616 \\
       \tabfmt{AC} & 3.852 & 3.351 & 2.661 & 1.467 & 1.650 & 0.624 \\
       \tabfmt{DC} & 35.064 & 42.892 & 47.428 & 23.134 & 1.202 & 0.454 \\
       \tabfmt{Hall} & 25.187 & 31.782 & 31.662 & 1.141 & 0.668 & 0.253 \\
       \tabfmt{QE} & 3.746 & 3.517 & 2.388 & 1.734 & 6.128 & 2.318 \\
       \tabfmt{WFT} & 3.446 & 3.091 & 2.330 & 1.658 & 6.184 & 2.339 \\
      \hline\hline
    \end{tabular}\end{table}

For a fair evaluation of the speedup of GPU versus CPU, normalization according to the FLOPS (floating-point operations per second) of the devices is required, since GPU and CPU may have different TFLOPS. The TFLOPS of A800 graphics card and Gold 6548Y+ CPU are 9.7 and 1.834 per device \cite{nvidia_a800, intel_6548}, yielding a normalization factor of $9.7/(1.834\cdot2)=2.644$. In the ideal situation, the normalized speedup should be approximately 1. As indicated by Table \ref{tab:tbpm_speedup}, LDOS, DOS, QE and WFT all have normalized speedup larger than 1, indicating that excellent GPU acceleration can be achieved. AC and DC have normalized speedup less than 1, possibly due to overhead arising from algorithmic complexity, memory access and GPU/CPU communication. DC and Hall conductivity have the lowest normalized speedup because they consume the most VRAM and have the largest overhead. Optimization of these algorithms is an important working direction of future development.
 
\section{Summary}
\label{summary}
In summary, we have introduced version 2.0 of \api{TBPLaS} package, a new major version that brings many improvements and new features for both users and developers. The Python/Cython modeling tools have been thoroughly optimized, and a new C++ version of the modeling tools has been implemented, enhancing the modeling efficiency by \todo{several orders}. The solvers have been rewritten in C++ from scratch following the philosophy of object-oriented programming and template meta-programming, leading to efficiency enhancement of \todo{several times or even an order of magnitude}. The workflow of using \api{TBPLaS} has also been unified into a more comprehensive and consistent manner. Other new features include spin texture, Berry curvature and Chern number calculation, partial diagonalization, analytical Hamiltonian, and GPU computing support. Documentation and tutorials have been updated. These new features and improvements not only enhance the efficiency and usability, but also improve the maintainability and extensibility of the package, making it an ideal platform for the development of advanced models and algorithms. Further developments and extensions, e.g.,  optimization of GPU version of TBPM algorithms, parallel TBPM algorithms on top of distributed sparse matrix library, support for single precision and multiple GPUs, transport properties calculation, the real-space self-consistent Hartree and Hubbard methods for large systems, will be implemented in the future.

~\\
{\textbf{CRediT author statement}}
~\\

\textbf{Yunhai Li}: Project administration, Software, Validation, Writing - Original Draft. \textbf{Zewen Wu}: Software, Validation, Writing - Original Draft. \textbf{Miao Zhang}: Software. \textbf{Junyi Wang}: Software. \textbf{Shengjun Yuan}: Conceptualization, Funding acquisition, Resources, Supervision, Writing - Review and Editing.

~\\
{\textbf{Acknowledgments}}
~\\

This work is supported by the National Natural Science Foundation of China (Grant NO. 12425407, NO. 12174291) and the Major Program (J.D.) of Hubei Province (Grant NO. 2023BAA020). Yunhai Li and Zewen Wu additionally acknowledge the National Natural Science Foundation of China (Grant NO. T2495255). Zewen Wu also acknowledges the Natural Science Foundation of Wuhan (Grant No.2024040801020388). The numerical simulations involved in this paper are performed on the supercomputer provided by Core Facility of Wuhan University.

~\\
{\textbf{Declaration of competing interest}}
~\\

The authors declare that they have no known competing financial interests or personal relationships that could have appeared to influence the work reported in this paper.

~\\
{\textbf{Data availability}}
~\\

Data will be made available on request.
 \appendix

\section{Installation notes}

\subsection{Preparation of virtual environment}
\label{virtual_env}

The virtual environment can be prepared using \api{pip}
\begin{lstlisting}[language=bash]
python -m venv tbplas $HOME/tbplas_install/tbplas
source $HOME/tbplas_install/tbplas/bin/activate
pip install numpy scipy matplotlib cython setuptools build
\end{lstlisting}
or alternatively using \api{conda}
\begin{lstlisting}[language=bash]
conda create -n tbplas python=3.12
conda activate tbplas
conda install numpy scipy matplotlib cython setuptools build
\end{lstlisting}

\subsection{Bash script for configuring environment variables}
\label{bmod}

The bash script which can be installed by
\begin{lstlisting}[language=bash]
# Replace VERSION_CPP with the actual version number
tar -xf tbplas-cpp-VERSION_CPP.tar.bz2
cp tbplas-cpp-VERSION_CPP/tools/init.sh $HOME/tbplas_install
source $HOME/tbplas_install/init.sh
echo "source $HOME/tbplas_install/init.sh" >> $HOME/.bashrc
\end{lstlisting}
Suppose we are going to build \api{TBPLaS} with OpenBLAS 0.3.28, HDF5 1.14.2 and FEAST 4.0. The dependencies have been built from source code and installed into the \api{\$HOME/tbplas\_install} directory, and the Python virtual environment we have prepared for the installation is named \api{tbplas}. Then the following bash commands will set up the relevant environment variables
\begin{lstlisting}[language=bash]
# HDF5
dest=$HOME/tbplas_install/hdf5-1.14.2
set_mod add pkg $dest
set_env add CMAKE_PREFIX_PATH $dest

# OpenBLAS
dest=$HOME/tbplas_install/openblas-0.3.28
set_mod add pkg $dest
set_env add CMAKE_PREFIX_PATH $dest

unset dest

# FEAST
reset_env add FEASTROOT $HOME/tbplas_install/FEAST/4.0

# Python environment
source $HOME/tbplas_install/tbplas/bin/activate
\end{lstlisting}
If the virtual environment has been created using \api{conda}, then the last line should be replaced with
\begin{lstlisting}[language=bash]
conda activate tbplas
\end{lstlisting}
Add the settings to \api{\$HOME/.bashrc} to make them permanently effective. If the dependencies have been installed from software repository, probably their paths are already included in the environment variables. In that case, skip the settings for HDF5, OpenBLAS and FEAST. Some dependencies may have their own instructions on setting up the environment variables, e.g., Intel oneAPI, AOCC, AOCL, CUDA toolkit and HPC SDK. Check the official installation guides of these dependencies for more details.

\section{Source code of auxiliary functions}

\subsection{cutoff\_pc}
\label{src:cutoff_pc}

The \api{cutoff\_pc} function is defined as following: firstly we get the Cartesian coordinates by calling \api{get\_orbital\_positions\_nm}, then loop over the coordinates to collect the indices of unwanted orbitals. Finally, we remove the orbitals with \api{remove\_orbitals} and trim dangling orbitals and hopping terms with \api{trim}
\begin{lstlisting}[language=c++]
void cutoff_pc(
    model_t& model,
    const Eigen::Vector3d& center,
    const double& radius = 3.0)
{
    std::set<size_t> idx_remove;
    Eigen::Matrix3Xd orb_pos = model.get_orbital_positions_nm();
    Eigen::Vector3d dr;
    for (size_t i = 0; i < orb_pos.cols(); ++i) {
        dr = orb_pos.col(i) - center;
        if (dr.norm() > radius) {
            idx_remove.insert(i);
        }
    }
    model.remove_orbitals(idx_remove);
    model.trim();
}
\end{lstlisting}

\subsection{extend\_hop}
\label{src:extend_hop}

The function \api{extend\_hop} is defined as following: firstly we obtain the hopping terms within cutoff distance with the \api{find\_neighbors} function, then we loop over the hopping terms and add them to the model. The hopping integrals are determined by the \api{calc\_hop} function following the Slater-Koster formulation \cite{PhysRevB.86.125413}
\begin{lstlisting}[language=c++]
double calc_hop(const Term& term)
{
    constexpr double a0 = 0.1418;
    constexpr double a1 = 0.3349;
    constexpr double r_c = 0.6140;
    constexpr double l_c = 0.0265;
    constexpr double gamma0 = 2.7;
    constexpr double gamma1 = 0.48;
    constexpr double decay = 22.18;
    constexpr double q_pi = decay * a0;
    constexpr double q_sigma = decay * a1;
    double dr = term.distance;
    double n = term.rij(2) / dr;
    double v_pp_pi = -gamma0 * exp(q_pi * (1 - dr / a0));
    double v_pp_sigma = gamma1 * exp(q_sigma * (1 - dr / a1));
    double fc = 1 / (1 + exp((dr - r_c) / l_c));
    double hop = (n * n * v_pp_sigma + (1 - n * n) * v_pp_pi) * fc;
    return hop;
}

void extend_hop(model_t& model, const double& max_distance = 0.75)
{
    auto neighbors = find_neighbors(model, model, 1, 1, 0, max_distance);
    for (const auto& n : neighbors) {
        cell_index_t ra, rb, rc;
        orbital_index_t orb_i, orb_j;
        std::tie(ra, rb, rc) = n.rn;
        std::tie(orb_i, orb_j) = n.pair;
        model.add_hopping(ra, rb, rc, orb_i, orb_j, calc_hop(n));
    }
}
\end{lstlisting}

\subsection{calc\_hop}
\label{src:calc_hop}

This is an overloaded version of \api{calc\_hop} which accepts an Eigen \api{Vector3d} reference as input. The other parts are the similar to that in \ref{src:extend_hop}
\begin{lstlisting}[language=c++]
double calc_hop(const Eigen::Vector3d& dr_vec)
{
    constexpr double a0 = 0.1418;
    constexpr double a1 = 0.3349;
    constexpr double r_c = 0.6140;
    constexpr double l_c = 0.0265;
    constexpr double gamma0 = 2.7;
    constexpr double gamma1 = 0.48;
    constexpr double decay = 22.18;
    constexpr double q_pi = decay * a0;
    constexpr double q_sigma = decay * a1;
    double dr = dr_vec.norm();
    double n = dr_vec[2] / dr;
    double v_pp_pi = -gamma0 * exp(q_pi * (1 - dr / a0));
    double v_pp_sigma = gamma1 * exp(q_sigma * (1 - dr / a1));
    double fc = 1 / (1 + exp((dr - r_c) / l_c));
    double hop = (n * n * v_pp_sigma + (1 - n * n) * v_pp_pi) * fc;
    return hop;
}
\end{lstlisting} 
\bibliographystyle{unsrt}
\bibliography{references.bib}

\end{document}